\documentclass[12pt]{iopart}
\usepackage[dvips]{graphicx}
\usepackage{amsfonts}

\begin{document}

\title{Self-pulsing effect in chaotic scattering}

\author{C.     Jung$^1$,   C.    Mej\'{\i}a-Monasterio$^1$\footnote[1]{Present
address: Center for Nonlinear  and Complex Systems, Universit\`a degli
Studi  dell'Insubria,  Via  Valleggio   11,  22100  Como,  Italy}, O. 
Merlo$^2$ 
 and T. H. Seligman$^1$}
\address{
$^1$Centro de  Ciencias   F\'{\i}sicas,  Universidad Nacional  Aut\'onoma  de M\'exico,  Cuernavaca,
Morelos, M\'exico \newline
$^2$ Institut f\"ur Physik der Universit\"at Basel, Basel 
Switzerland}

\ead{carlos.mejia@uninsubira.it}

\date{\today}

\begin{abstract}
We study the  quantum and classical  scattering of Hamiltonian systems
whose chaotic saddle is described by  binary or ternary horseshoes. We
are interested in parameters of the system  for which a stable island,
associated  with  the inner  fundamental  periodic orbit of the system
exists and is large, but  chaos around this  island is well developed.
In  this situation, in  classical systems, decay  from the interaction
region is algebraic, while in quantum systems it is exponential due to
tunneling. In  both cases, the  most  surprising effect is  a periodic
response to  an incoming wave packet. The  period of this self-pulsing
effect  or scattering echoes coincides with  the mean period, by which
the scattering  trajectories rotate   around the stable  orbit.   This
period of rotation is directly related to the development stage of the
underlying horseshoe.    Therefore the  predicted  echoes will provide
experimental access to  topological information.   We numerically test
these results in kicked one dimensional models and in open billiards.
\end{abstract}

\submitto{\NJP}

\maketitle

\section{Introduction}
\label{sec:intro}

The purpose of any  scattering experiment is  to learn something about
the  dynamics of the total system  consisting of projectile and target
in  situations, where the direct observation  of  the dynamics of this
total system  is not possible or  difficult. Then  we need to evaluate
the asymptotic scattering data to gain  this desired insight.  Thereby
we are   lead to the inverse  scattering  problem.  Traditionally this
expression is used for  the reconstruction of the  potential or of the
Hamiltonian. For chaotic systems, however, the reconstruction of other
quantities  like for example  characteristics of the chaotic saddle or
chaotic invariant  set  might also be worth  while  and more realistic
\cite{lipp99}.  Even if possible,  reconstructing the  potential first
and using  it to reconstruct the dynamics  and then investigating this
reconstructed  dynamics to get the information  on  the chaotic saddle
includes a lot of sources for  errors and ambiguities. Therefore it is
attractive   to develop strategies  to  extract  the properties of the
chaotic saddle directly from the  scattering data.  In this spirit the
most complete information would be  the reconstruction of the complete
knowledge  of  all properties  of the  chaotic   saddle of  the system
directly from  asymptotic measurements.  For  system of two degrees of
freedom (including time-dependent systems  with one degree of freedom)
this chaotic set might be   represented by a horseshoe  \cite{smale67}
construction of  an iterated map,  typically  on a surface  of section
appropriately  chosen.  The inverse problem turns  into the problem of
reconstructing the properties of this  horseshoe from the knowledge of
scattering data.

First steps  in this  direction  have been taken  in \cite{lipp99} for
kicked systems in one  dimension as well as  for systems with one open
degree of  freedom and   one   closed vibrational  degree of   freedom
\cite{tapia03}.   An independent   alternative has  been  presented in
\cite{buetikofer00}.  All this work is  based on the evaluation of the
fractal structure of scattering functions where the hardest problem is
the recognition   of  the  hierarchical  level  of all   intervals  of
continuity.  Once  we have done   this  step we  can reconstruct   the
branching tree of   the underlying fractal,  measure  all its  scaling
factors and apply the  thermodynamical methods \cite{tel90} to extract
quantitative    measures  of chaos.  For     the measurement of  these
scattering functions the simultaneous  preparation and  measurement of
canonically  conjugate  variables  is necessary.  Therefore all  these
methods are essentially  classical and might at best  work in the deep
semi-classical domain  of wave mechanics.   In addition  these methods
are useful only for complete or near-complete horseshoes.

In contrast in the  present paper  we  present a  completely different
idea which is  based on the  self-pulsing effect of chaotic scattering
systems with horseshoes of a low development stage. The present method
has  the pleasant property that  it does  not require the simultaneous
measurement of conjugate   variables. Therefore it works  similarly in
classical  and in  quantum  dynamics.  The only important   difference
between the classical and the wave versions is,  that due to tunneling
in wave mechanics the interior of KAM islands  is also explored, which
is inaccessible to asymptotic observation  in the classical case.  For
our argument it is  essential that the  system can be described by  an
iterated   map  on a two-dimensional  domain.   This implies  the same
restriction   to two dimensions,   that   we encountered  in  previous
work\cite{lipp99,tapia03,buetikofer00}.

Preliminary  results along for the line   of argument we shall present
were given  in a letter  \cite{mejia01} for kicked systems. At present
experiments are   in progress in   Darmstadt \cite{exp04} applying our
idea to   the  propagation  of   electromagnetic  waves  in   an  open
superconducting cavity that can  be viewed as an open  two-dimensional
billiard.

We  shall  present  both  kicked  systems  for  their  simplicity  and
scattering billiards because of the ongoing experiments and because of
their relevance for mesoscopic structures. The surface of section will
be  a stroboscopic map  in the  first case  and a  Birkoff map  in the
second one. We shall concentrate  on geometries that lead to low order
horseshoes, in  fact limiting our  considerations to second  and third
order,  though the basic  idea is  not subject  to such  a limitation.
Furthermore we shall make extensive  use of the parametrization of the
development of a horseshoe as given in \cite{rueckerl94}, and we shall
only be concerned by  degrees of development considerably smaller than
one half, which  insures, that a large stable  island around the inner
fixed point  still exists.  We shall show,  that this island  with its
characteristic fractal  surroundings rotate at  a rate defined  by the
formal development parameter of the horseshoe.  This gives rise to the
suspicion, that  this time of  revolution can be measured  by studying
pulsed reverberance of the scattering system, which we name scattering
echoes  (to  distinguish   them  from  Loschmidt  echoes).   Numerical
simulation shows, that it should be possible to measure such echoes.

In the  next section we  shall proceed to  discuss the way such echoes
may  occur, and in the  third section we  derive the  formulae for the
relation between formal   development parameters of the horseshoe  and
the period of the echoes. We then proceed  to discuss examples leading
to binary  and ternary horseshoes. For  kicked systems we discuss both
the classical and wave-mechanical part,  while we limit ourselves to a
classical discussion of scattering billiards,  as these present a fine
example where reliable  calculations are extremely expensive, and will
not  be able to replace  the experiment in  the  microwave cavity.  We
will then draw some conclusions and give an outlook.

\section{Scattering Echoes}
\label{sec:echoes}

In this section, we shall demonstrate  the appearance of echoes in the
scattering    of any   typical  two  degrees   of freedom  Hamiltonian
scattering system described  by a binary  or ternary  horseshoe with a
low degree  of development, {\it  i. e. } with  a large central stable
island.

We  first    consider a  scattering  system  whose  chaotic  saddle is
described  in some appropriate Poincar\'e  map  by a binary horseshoe,
i.e. one having  two fixed points.  One  of the fixed points is normal
hyperbolic for any set  of values of the   physical parameters of  the
system.  It couples  the complicated  dynamics inside the  interaction
region  with the regular dynamics in  the asymptotic region.  For this
reason, this unstable fixed point is  usually called the {\it exterior
fixed point}.  Its invariant manifolds  trace out the horseshoe,  from
whose topological structure  at least an approximate symbolic dynamics
can be obtained \cite{lipp99,rueckerl94}.  This forms  the core of the
method presented in this paper.

The intersections between  the  stable and  unstable manifolds of  the
horseshoe  form the  so-called   homoclinic tangle.   The  fractal and
hierarchical    arrangement of   the    homoclinic   intersections are
responsible for the  chaotic motion inside  the scattering region  and
surrounds the second  fixed point of the  Poincar\'e map,  thus called
the {\it  inner fixed point}.  Contrary  to  the exterior fixed point,
the stability of the   inner fixed point   depends  on the  values  of
physical  parameters  of   the system.     For  typical  systems,  the
Poincar\'e  map shows a complete KAM   scenario: First the inner fixed
point is stable, surrounded   by KAM surfaces.  Eventually,  those KAM
surfaces  break into chains  of  secondary islands and chaotic  layers
which for higher levels of development  fuse into one large homoclinic
tangle.  For some   values of the  parameters, the  inner fixed  point
changes its stability through a period-doubling bifurcation.

\begin{figure}[!t]
\begin{center}
\includegraphics[scale=0.5]{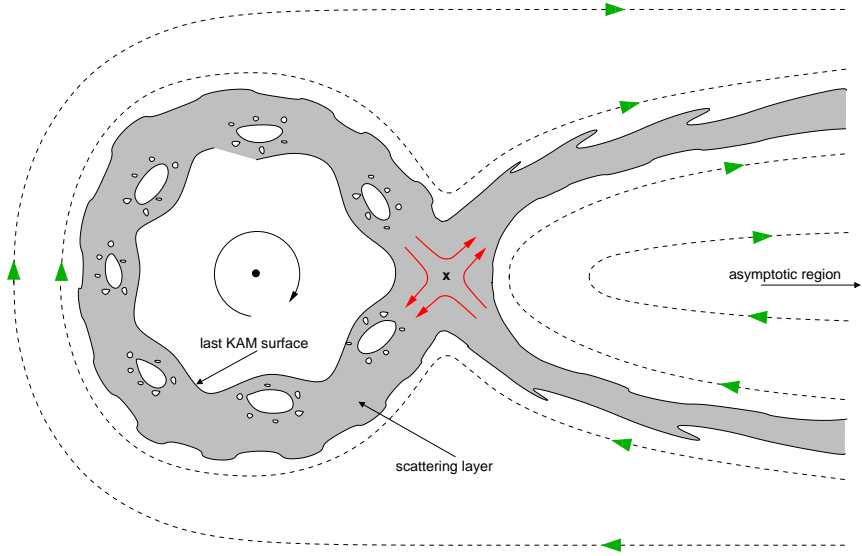}
\caption{\label{fig:sos-2}
Schematic   representation  of the  generic  surface  of section for a
scattering  system described by a binary   horseshoe.  The green arrow
lines  represent the  flow  of the  trajectories  over  the surface of
section.}
\end{center}
\end{figure}

For the method presented here the  existence of an inner stable island
is  essential and   therefore,  we shall focus  on   a region  in  the
parameter space  for  which  the inner   fixed  point is   stable  and
surrounded  by a large  primary  island of  stability with the generic
region of  secondary  stable  islands and chaos   surrounding  it.  In
Fig.~\ref{fig:sos-2}   we  show  a  schematic   representation of  the
structure of the  Poincar\'e map of  a generic system in the situation
just described. The exterior fixed point is represented by a cross and
its unstable dynamics in its vicinity  is indicated by the red arrows.
The homoclinic tangle of the exterior fixed point surrounds the stable
island.    Beyond  the last  KAM  surface of   the  primary island the
non-trivial scattering trajectories traverse what we may call the {\it
chaotic scattering layer}.

The scattering layer extends to infinity, covering the whole domain of
the map  outside of the last  KAM  surface. However,  the  part of the
scattering layer of interest for the scattering process corresponds to
the one  whose energy domain in  the  asymptotic region coincides with
the    energy domain for which   the  singular structure in scattering
functions  appear.  This chaotic   scattering  layer is  actually  the
region of  the homoclinic tangle  represented in Fig.~\ref{fig:sos-2},
as the  gray  ring around  the stable  island.   The gray  strips that
extend to  infinity correspond  to region  in which  the  incoming and
outgoing trajectories enter or leave  the chaotic scattering layer and
are the region over which the  infinity of tendrils  of the stable and
unstable manifolds of the outer fixed  point are spread.  The incoming
trajectories not belonging to this gray  layer will either bounce back
to   the asymptotic region   immediately or  travel around the  stable
island without  entering  into  the chaotic  scattering layer.   Those
trajectories  will  be scattered   in  what are usually called  direct
processes.

Let us describe the  scattering  process as  viewed on the  Poincar\'e
surface  of  section  sketched in Fig.~\ref{fig:sos-2}.    We focus on
incoming   scattering trajectories  that   start  close to  the stable
manifold.  When these trajectories  enter the chaotic scattering layer
they rotate around the  stable island.  This rotation is characterized
by the  winding number of the outermost  KAM surface.  Its effects are
felt over the full  scattering layer and as  we shall show in the next
section, it can be directly related  to the topology of the underlying
chaotic saddle.

As is  well known \cite{moser73}, in their  rotation around the island
the trajectories  always return to the  vicinity of the exterior fixed
point.  Depending on the initial conditions, after one revolution some
of  the trajectories   will  leave the  homoclinic  tangle  along  the
unstable   manifold.   The rest will  continue    in their rotation to
complete a second  revolution around the  island.  When the  remaining
trajectories return to the vicinity  of the exterior fixed point, part
of    them will in  turn exit   the  chaotic  scattering  layer to the
asymptotic region. This process will continue {\it ad infinitum}.

The key observation is the following: Any trajectory starting close to
the  stable manifold will leave  the chaotic scattering layer after an
integer number of revolutions around the  stable island.  As a result,
if we measure the  asymptotic  outgoing flux  resulting from a  narrow
packet  of incoming trajectories,   we   shall observe that   the flux
intensity oscillates in time. Clearly, the time between two successive
maxima of  the outgoing flux  will correspond to the  period of time a
typical trajectory  in    the scattering layer   needs   to complete a
rotation around the island or equivalently,  to the return time to the
neighbourhood of  the exterior fixed point.   The minimal  return time
corresponds to the  trajectories close to  the  surface of the  island
which is  most sticky  and  thereby, determines the  time  between the
observed  outgoing pulses, {\it i.e.}   the  scattering echoes.  In  a
typical situation the return times  of the trajectories in the chaotic
scattering layer  increase  monotonously with   the distance from  the
center of the stable island.   Therefore, the time between  successive
echoes will be determined  by the mean  orbital period of rotation  of
the trajectories rotating close  to the  surface of  the island.   The
width  of  the echoes in   the outgoing flux  will   be related to the
intrinsic  dispersion of the  packet  of incoming trajectories in  the
chaotic scattering layer.   Thus, the appearance  of scattering echoes
is typical for  $2$-dimensional Hamiltonian scattering  systems, where
the   chaotic  invariant set  is represented    by a horseshoe  of low
development stage.

\begin{figure}[!t]
\begin{flushright}
\includegraphics[scale=0.5]{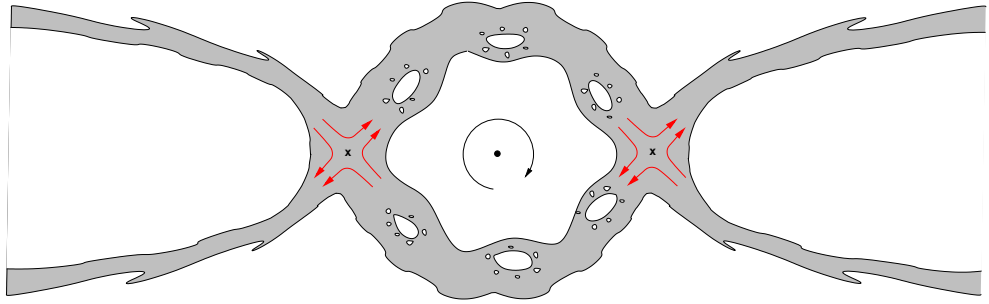}
\caption{\label{fig:sos-3}
Schematic representation  of the  generic   surface of section  for  a
scattering system described by a ternary horseshoe.}
\end{flushright}
\end{figure}

Hamiltonian scattering systems in which the Poincar\'e map possesses 3
fixed points can be described by a ternary horseshoes.  We concentrate
on   cases where  there  exist two  exterior   normal hyperbolic fixed
points.  This implies the existence of two fractal bundles of tendrils
of stable  manifolds  along which trajectories  can  enter the chaotic
layer and another two fractal consisting of unstable manifolds for the
exit. We shall continue to call the tangle formed by the intersections
of  the invariant  manifolds   of   the  two  exterior  fixed   points
"homoclinic  tangle", though it  clearly contains  both homoclinic and
heteroclinic intersections.  It  surrounds the inner fixed point whose
stability is determined by the physical parameters  of the system.  In
Fig.~\ref{fig:sos-3} we   show  a   schematic representation of    the
structure  of the Poincar\'e  map of a generic  system described by a
symmetric ternary horseshoe.   In the region  of parameter space where
the inner fixed  point  gives  rise  to a  well developed  island   of
stability surrounded by a chaotic scattering  layer, it is possible to
follow the  same  line of  argumentation  as above. This  leads to the
conclusion that the existence   of scattering echoes remains  generic.
The time   between echoes is  again given  by the return   time in the
chaotic scattering layer to the given  exterior fixed point.  The main
difference with the binary case is that for the symmetric ternary case
there are two outgoing manifolds near to which the self-pulsing can be
measured.  If we  perform such an experiment with  a  narrow packet of
incoming trajectories from one side the scattering echoes shall appear
in both outgoing manifolds in counter-phase.

On  the quantum  level we expect  to  see the same  scattering echoes.
Actually, we  find that the echoes  are more clean-cut  in the quantum
regime,  and  we  shall see that   this is  due  to tunneling.   While
forbidden  for  classical  trajectories the  quantum   probability can
penetrate KAM  surfaces and settle on inner  regions of  the classical
island.    Therefore, if the quantum    probability rotates around the
island at a  smaller   distance from the  center,  the  period of  the
quantum echoes will systematically differ from  the classical one. The
quantum probability will feel the winding number of inner KAM surfaces
and send this  information out  through the  scattering echoes.   This
phenomenon is interesting  by itself.  By measuring  the period of the
quantum echoes we  are able to probe regions  in phase  space that are
inaccessible to classical  scattering and obtain information about the
topology of  the classical chaotic  saddle  by controlling  the  limit
$\hbar\rightarrow 0$.
\section{The period of scattering echoes}
\label{sec:method}

The central  point  of  the method  developed  in  this paper   is the
connection   between  the period  of  the scattering   echoes  and the
development stage of the   horseshoe.  As has  been discussed  in  the
previous  section, this time corresponds  to  the inverse  of the mean
winding number or orbital period in the chaotic scattering layer. Thus
the period of the  scattering  echoes is given   in units of the  mean
return time to the surface of section.  To unravel its connection with
the topology of the underlying chaotic invariant set  we shall use the
formal development parameter $\alpha$  as a measure of the development
of the horseshoe.   It measures the penetration  depth of the tendrils
into the  interior of  the  fundamental  area  of  the  horseshoe  and
therefore contains the information      on the  primary     homoclinic
intersections. This tells us the  universal aspects of the topology of
the hyperbolic component of the horseshoe.  The parameter takes values
from $0$ corresponding to a separatrix loop, to $1$ when the horseshoe
becomes complete.   For   a  detailed definition of  the   development
parameter  and a  discussion of the   development stages of incomplete
horseshoes as well as for the terminology we use,  we refer the reader
to Refs.~\cite{lipp99,rueckerl94}.

The  relation of  the  development parameter  $\alpha$ to the  winding
number in  the chaotic scattering layer is  most easily established if
$\alpha=1/k^n$, where   $k$  is the  order   of the  horseshoe.  Thus,
without  loss of generality, we  shall  stick to  this  case and  then
invoke continuity arguments to close the gaps.

\begin{figure}[!t]
\begin{center}
\includegraphics[scale=0.6]{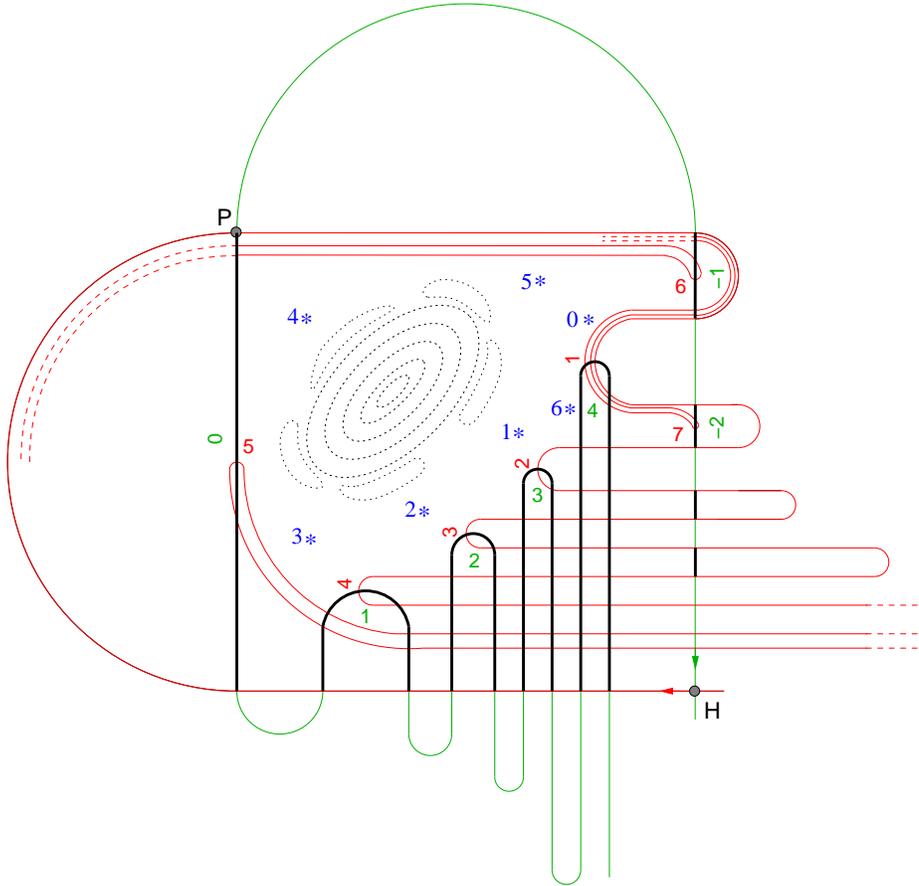}
\caption{\label{fig:method}
Topology of   a  horseshoe for  $\alpha=1/16$.  The  point  $H$ is the
exterior fixed point.  Its invariant manifolds the unstable in red and
stable in green,   first intersect in  the  primary intersection point
$P$.  The successive  tendrils  are labeled   starting  from $P$.  The
tendrils of the stable manifold have been emphasized in black.}
\end{center}
\end{figure}

Before continuing with  the description of the  method, there are some
properties   of      the   horseshoe  construction,    illustrated  in
Fig.~\ref{fig:method}, we would like to point out:

\begin{enumerate}
\item The horseshoe  construction   is   created   by  the   intricate
intersections between    the stable  and  unstable manifolds    of the
exterior  fixed points.  The horseshoe exists  in a surface of section
and its outer  unstable part forms  the chaotic scattering  layer. The
invariant manifolds  of the exterior fixed points  never enter any KAM
island.

\item Any point on the unstable invariant manifold  converges  to  the
exterior fixed point as time $t\to - \infty$.  Any point on the stable
invariant manifold   converges to the   exterior fixed  point  as time
$t\to\infty$.

\item The labeling of the different tendrils of the horseshoe is quite
convenient as due to the invariance  of the manifolds, each tendril is
the image  and the pre-image  of  other tendrils under  the Poincar\'e
map. For the unstable manifold tendril $n$ is the image of $n-1$.  For
the stable manifold tendril $n$ is the image of $n+1$.
\end{enumerate}

In what  follows  we shall  stick  to the  binary horseshoe   case and
briefly discuss at the  end the modifications of   the method for  the
ternary    case.    By definition,    if     for a    binary horseshoe
$\alpha=2^{-n}$, the  first tendril of  one of the invariant manifolds
is intersected by  the tip of  the $n$-th tendril  of the other one in
two points  only.  We label the different  tendrils  starting from the
so-called primary intersection indicated by the  letter $\mathsf P$ in
Fig.~\ref{fig:method}.  There  we   show  schematically the   topology
corresponding to  an incomplete  horseshoe with $\alpha=2^{-4}$. Thus,
tendril $1$  of  the  unstable  invariant manifold (shown  in  red) is
intersected by  the tip of the $4$-th  tendril of the  stable manifold
(in green) in two points.

In Fig.~\ref{fig:method}, we  have  emphasized  the tendrils of    the
stable manifold in  black. From property (iii)  the tendril $1$ of the
stable manifold  is the image of  tendril $2$ and pre-image of tendril
$0$.   Tendril $2$ is in  turn the image  of tendril $3$ while tendril
$0$  is the pre-image  of  tendril $-1$  and so   on back and  forward
{\it ad infinitum}.  Due to  this hierarchy, an  intersection between
tendril  $n$ of the unstable manifold   and tendril $m$  of the stable
manifold  labeled by $(n\!:\!m)$  is  the  image  of the  intersection
$(n\!-\!\!1\!:\!m\!\!+\!\!1)$            and     pre-image          of
$(n\!+\!\!1\!:\!m\!\!-\!\!1)$.

We now  have all the information we  need to relate the development of
the  horseshoe   in terms of   the parameter  $\alpha$  to the orbital
period.   By    construction, tendril $1$  of     any of the invariant
manifolds is in between tendrils $-1$ and $-2$  of the other manifold.
If a generic trajectory  is started  in  the vicinity of  the unstable
tendril   $1$     (represented  by  the      sequence   of  stars   in
Fig.~\ref{fig:method}),  after one iteration  of the Poincar\'e map it
will be in the vicinity  of the unstable tendril  $2$.  After a second
iteration, the trajectory will be near the unstable tendril $3$ and so
on. {\it A priori}, we do not know which of the unstable tendrils will
be again in the  vicinity of the unstable  tendril $1$, but we do know
that this trajectory will  have  almost completed a revolution  around
the inner stable  island when it   reaches the vicinity of  the stable
tendril  $-1$,  and a little  bit  more  than  one  revolution when it
reaches  the vicinity of   the  stable  tendril $-2$.  A   development
$\alpha=2^{-n}$ implies an intersection $(1\!:\!n)$.  Since the stable
tendril $-1$ is the ($n\!+\!1$)-th  image of  the stable tendril  $n$,
the unstable tendril  which intersects the  stable tendril $-1$ is the
($n\!+\!2$)-th.  Thus, if the  trajectory was started in  the vicinity
of the unstable tendril $1$,  and since the unstable tendril $n\!+\!2$
is the ($n\!+\!1$)-th image of  $1$, this trajectory will have  almost
completed a revolution in  $n\!+\!1$ iterations of the Poincar\'e map.
Equivalently, this trajectory  will have completed  a  little bit more
than one revolution after $n\!+\!2$ iterations of the Poincar\'e map.

Concerning our interest  in the average  behaviour of the trajectories
close to the surface of the island, we have obtained the following: If
the  development of  the underlying horseshoe  is $\alpha=2^{-n}$, the
orbital period    of  the  scattering  trajectories is   approximately
$n\!+\!3/2$ and we cast this result into the form
\begin{equation} \label{eq:magic_2}
T = -\log_2 \alpha + \frac{3}{2} \ .
\end{equation}

The mean period  $T$ in eq.~\ref{eq:magic_2} is given  in units of the
return time  to the surface of section  $\tau_{SOS}$ which corresponds
to the  time spend in one  iteration of the Poincar\'e  map.  For time
periodic systems, the Poincar\'e map is the stroboscopic map and thus,
the  return time  to  the surface  of  the section  is constant.   For
systems which are not periodic in  time, the units of $T$ can be taken
as  the mean  return time  to the  surface of  section in  the chaotic
scattering layer.   The development of  the horseshoe, related  to the
dynamics of the  system, does not depend on the  surface of section we
select.  This  is tantamount  to the fact  that the  mean $\tau_{SOS}$
does not depend on the surface we choose.

In  the derivation  of   eq.~\ref{eq:magic_2}  we  have  not  used any
particular model.   Therefore, this relation  between the mean orbital
period and the  topology of the hyperbolic  component of the horseshoe
given in terms of the  development parameter $\alpha$ is universal for
Hamiltonian scattering   systems   described by  a   binary horseshoe.
Inverting eq.~\ref{eq:magic_2} we obtain
\begin{equation} \label{eq:inv_magic_2}
\alpha(T) = 2^{(\frac{3}{2}-T)} \ .
\end{equation}
However, since  $\alpha$ is given in terms  of  the revolutions of the
scattering trajectories around the stable  island, the application  of
eqs.~\ref{eq:magic_2} and  \ref{eq:inv_magic_2}   is  limited to   the
situation   where a central     stable  island actually exist.    This
condition is formally fulfilled for  $\alpha < 1/2$ \cite{rueckerl94},
but practically this method is probably limited to  cases in which the
horseshoe is in a stage of low development say with $\alpha
\approx 1/4 $ or smaller.

For    the  case of  Hamiltonian  scattering  systems   described by a
symmetric ternary horseshoe, following  the same arguments as for  the
binary case,  the  formal  development parameter (called    $\beta$ to
distinguish   it from  the binary case)   is  related with the orbital
period by
\begin{equation} \label{eq:inv_magic_3}
\beta(T) = 3^{(\frac{3}{2}-T)/2} \ .
\end{equation}
In eq.~\ref{eq:inv_magic_3}, the base $3$ comes from the fact that the
horseshoe is  ternary and  the  $1/2$ in the  exponent  is due to  the
existence of two exterior fixed points.

\section{Examples}
\label{sec:examples}

In this  section we shall consider both  binary and  symmetric ternary
situations in detail using   kicked one-dimensional systems and   open
two-dimensional  billiards   as  examples.    On  one  hand  we  shall
concentrate on a billiard that gives rise to a binary horseshoe, while
another one corresponding to a ternary horseshoe  will be presented in
the  context  of a  superconducting experimental realization performed
presently at Darmstadt \cite{exp04}.  On the other hand we shall study
in detail  a kicked system  that gives  rise  to  a ternary  horseshoe
giving  considerably more detail than in  the  short discussion of the
binary horseshoe in Ref.~\cite{mejia01}.  For kicked systems classical
and quantum dynamics are  easily solved numerically. For billiards the
classical analysis is simple, while the numerical analysis of the wave
equation  in scattering   situations  is  more  delicate,   and little
previous work has been done.  Here we shall limit our analysis of such
systems to the classical case and leave the study of wave mechanics to
the experimental analysis.

\subsection{Binary horseshoe}
\label{sec:binary-horseshoe}

\begin{figure}[!t]
\begin{center}
\includegraphics[scale=0.6]{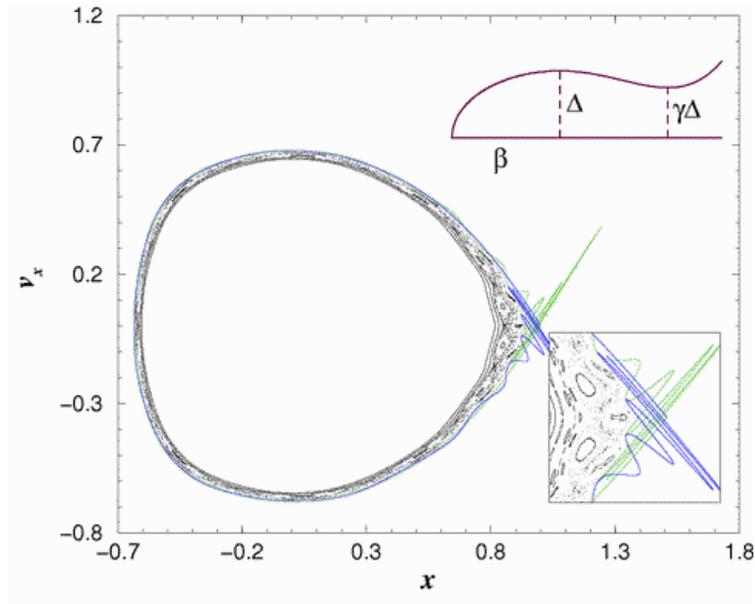}
\caption{\label{fig:bottle-phase-portrait}
Classical phase portrait for   the bottle billiard  for $\Delta=0.35$,
$\beta=1$  and   $\gamma=0.75$.     The horseshoe   with   development
$-\log_2\alpha=11$ is shown  in   blue  (stable manifold) and    green
(unstable manifold).  The  shape of the billiard  is shown at  the top
right of the figure  where the dashed  lines corresponds to the stable
(interior) and unstable (exterior)  periodic orbits.  In the inset  we
show an amplification or the region around the exterior fixed point.}
\end{center}
\end{figure}

In this section we shall study the classical echoes that appear in the
scattering of  a classical open  elastic billiard.  For this  we could
think of  using the well  known three discs  problem \cite{threedisk},
but this system  has serious disadvantages; the principal  one is that
the low development degree of  the horseshoe, that we desire cannot be
reached.  While the shadowing of orbits from one disc to the next will
lead to incomplete horseshoe  scenarios due to pruning, stable islands
do not develop.  This can be intuitively understood from the fact that
even when the circles osculate such islands are absent \cite{islands},
and a numerical analysis confirms  that the horseshoes we need are not
obtainable.  We  therefore look  for a bottle  shaped billiard  with a
single opening.  Its boundary consist in an arc of ellipse to the left
(bottom  of  the  bottle)  and  a quartic  polynomial  curve  properly
connected (for the neck of the bottle).  For practical purposes we use
only half the bottle split  along its length.  Therefore, at the lower
part the billiard will consist in the line $y(x) = 0$.  The exact form
of the upper boundary is
\begin{equation} \label{eq:bottle}
y(x) = \Delta\left\{
\begin{array}{ccl}
(1-x^2/{\beta^2})^{1/2} & , & x<0 \\
& & \\
(\frac{x^4}{16\beta^4(1-\gamma)} -\frac{x^2}{2\beta^2}+1)& , & 
x\ge 0 \\
\end{array}
\right..
\end{equation}
The      shape     of     the      bottle     is      shown     inside
Fig.~\ref{fig:bottle-phase-portrait}   where  the   length  parameters
$\beta$, $\gamma$ and $\Delta$  are indicated. The parameters $\Delta$
and $\beta$  correspond to the  semi-axes of the ellipse;  $\gamma \in
[0,1)$  corresponds   to  the  size   of  the  opening  in   units  of
$\Delta$. This  boundary is a ${\mathcal C}^2$  function everywhere in
its domain.

The  bottle has  two   fundamental  periodic orbits   corresponding to
trajectories, which are normal to both the upper and lower boundaries.
These       orbits    are      shown     as   dashed         lines  in
Fig.~\ref{fig:bottle-phase-portrait}.  The  stability of    the  inner
periodic  orbit  at  $x=0$   depends on  the local    curvature of the
boundaries  and therefore, on the values  of all  the parameters.  The
exterior periodic orbit at  $x_n = 2\beta\sqrt{1-\gamma}$ is  unstable
for  any set  of parameters.   It couples  the   relevant part of  the
complicate dynamics in  the interior  of the  bottle with the  regular
dynamics in   the asymptotic region.  In our   surface  of section the
exterior periodic   orbit  will  give a    hyperbolic fixed  point  at
$(x,v_x)=(x_n,0)$.   Its  stable   and  unstable  invariant  manifolds
constitute the horseshoe that for this billiard is binary.  Therefore,
by tuning the different parameters we can study the scattering process
for   different  stages   of  development   of   the  horseshoe.    In
Fig.~\ref{fig:bottle-phase-portrait} a     Poincar\'e     section  for
$\Delta=0.35$, $\beta=1$ and  $\gamma= 0.75$ is  plotted.  It shows  a
large stable island  surrounded by a  well  developed scattering layer
over  which  the corresponding horseshoe (in  blue  and green) is also
shown.     Each point in  the   surface of  section  correspond to the
position  $x$  and $x$-component of  the velocity  $v_x$ of a particle
trajectory when it hits  the lower ($y=0$)  boundary of the  billiard.
For this case, the development of the horseshoe is $-\log_2\alpha=11$.

\begin{figure}[!t]
\begin{center}
\includegraphics[scale=0.45]{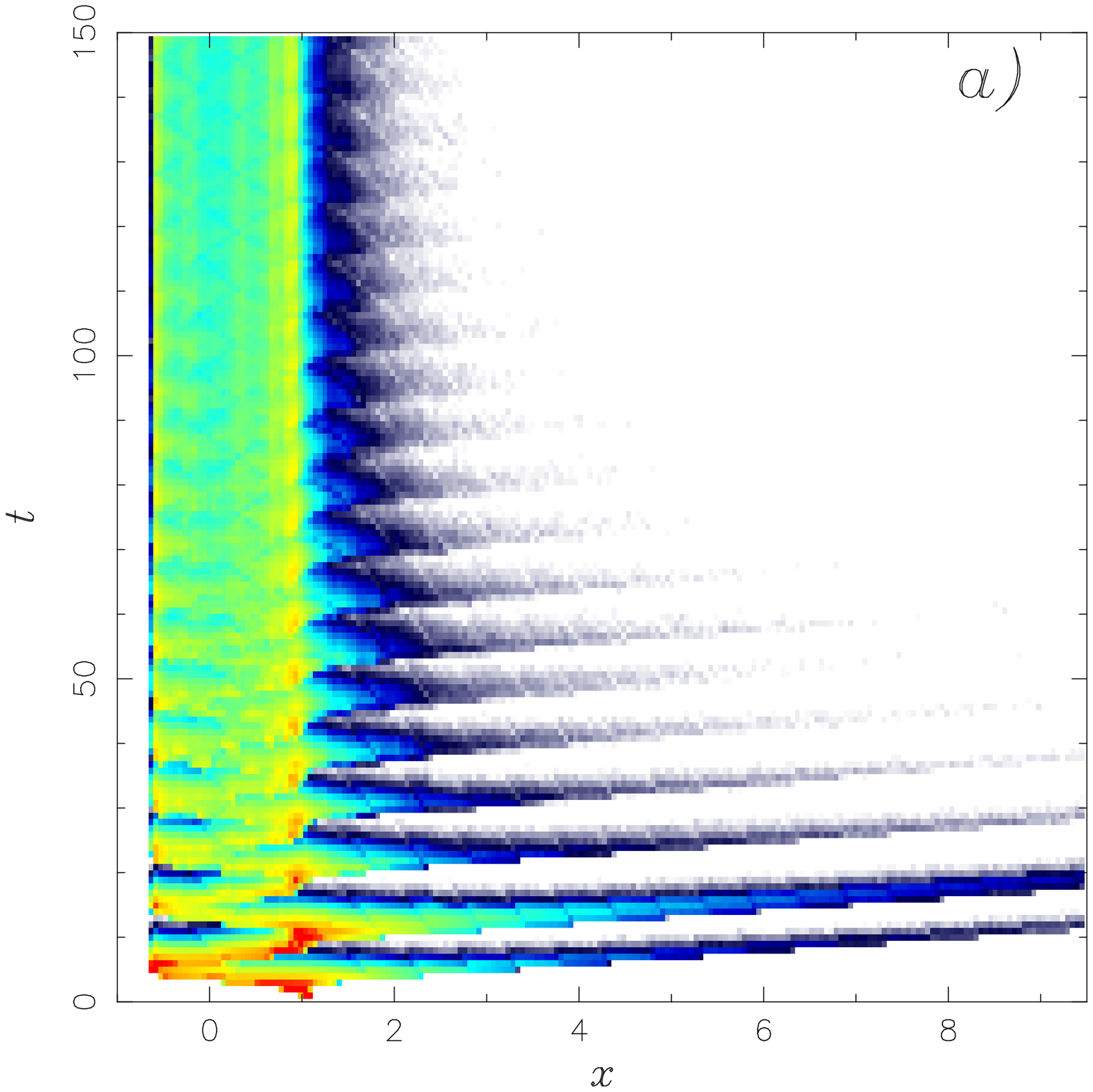}
\includegraphics[scale=0.4]{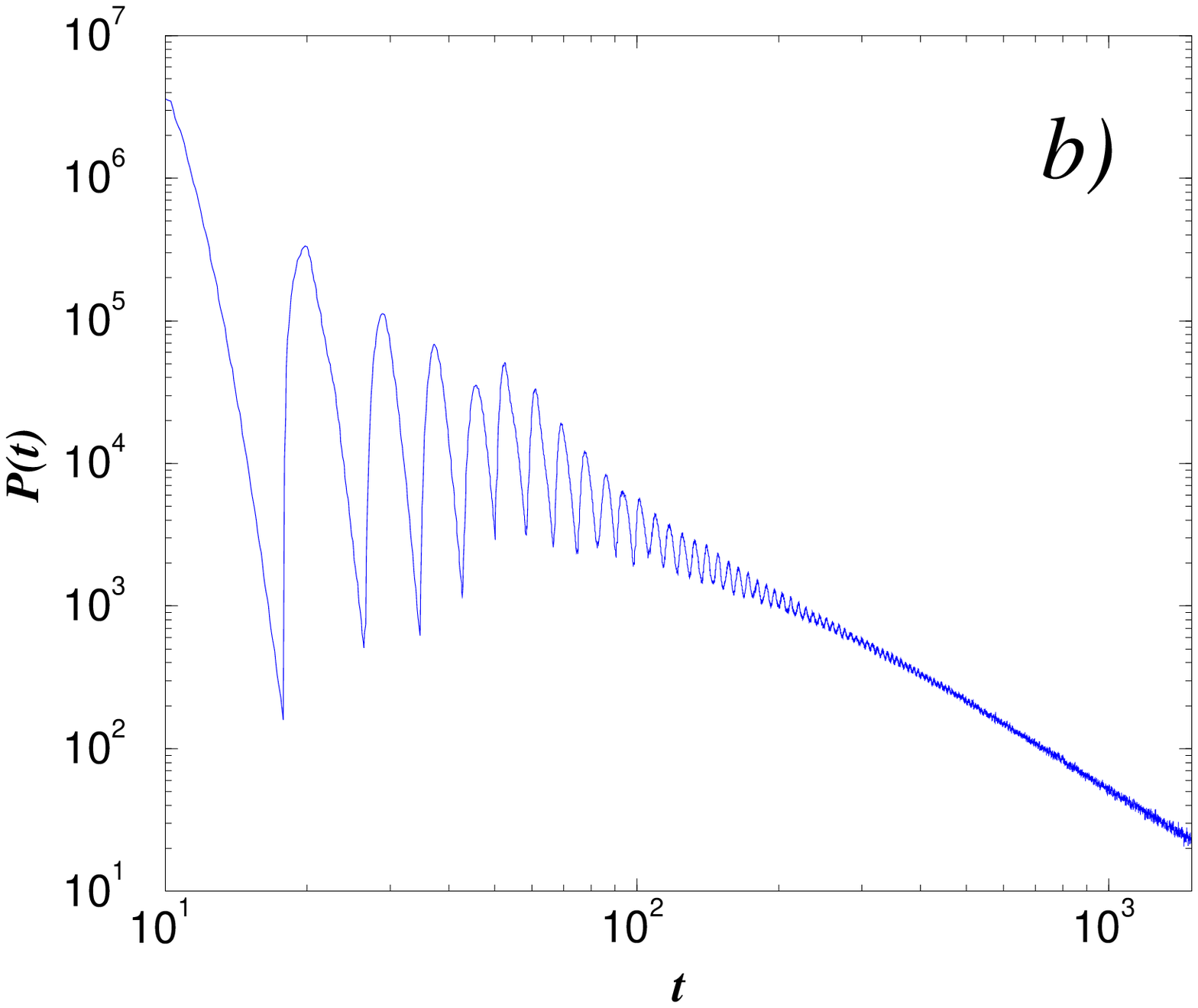}
\caption{\label{fig:bottle-echoes}
$a$) Classical density distribution in space and time $\rho_{\rm cl}(x,t)$
for the bottle  billiard  in a logarithmic  colour scale  for the same
parameters  as   in  Fig.~\ref{fig:bottle-phase-portrait}.  The colour
ramp increases from blue to red.  White  indicates very low densities.
$b$) Outgoing flux  of classical trajectories  as  a function  of time
$P(t)$ measured at $x=20$.}
\end{center}
\end{figure}

To study  classical scattering    in  this  model system,   we    have
numerically solved  the    classical  ray    dynamics of    scattering
trajectories inside the billiard.  To observe the scattering echoes we
have  followed    the   {\it    gedanken}  experiment    proposed   in
Sec.~\ref{sec:echoes}: We have  prepared a  narrow packet  of  initial
conditions in phase space.  Initially, the  narrow packet is placed in
the asymptotic  region.  Any  trajectory initially  not close  to  the
incoming  stable manifold of the horseshoe  will  either not enter the
chaotic  scattering layer or  will enter  but  leave immediately in an
uninteresting direct  process.    Since  we are  interested  in  those
trajectories  that enter the  scattering  layer the initial velocities
$v_x$ were chosen  randomly in a  small window around some appropriate
value,   such that  the initial  packet   shadows the incoming  stable
manifold of the   horseshoe.  As ray  dynamics   trivially scales with
energy, we fix the speed of all trajectories to one.  Therefore, $v_x$
can be interpreted as  $\cos{\phi}$ where $\phi$  is the angle between
the direction of the trajectory and the normal to  the boundary in the
position  of collision.  In  Fig.~\ref{fig:bottle-echoes}-$a$, we show
the classical density  $\rho_{\rm cl}(x,t)$ as a  function of position and
time for $\Delta=0.35$,  $\beta=1$ and $\gamma=  0.75$.  Following the
scattering process   we     can observe the    initial  narrow  packet
approaching  the  scattering region from  the  right.  When the packet
reaches the bottle neck at $x_n=1$ some  trajectories are bounced back
while the rest  enter the  bottle.  On  the surface  of section, these
trajectories rotate in the scattering layer, around the stable island.
This rotation is clearly seen as  an oscillation in the $x$-$t$ plane.
After  one oscillation  (corresponding  to one   revolution around the
stable island) some of  the   trajectories exit the  scattering  layer
along the  unstable manifold.  The rest  are bounced back at  the neck
staying in the scattering layer one  more oscillation.  After a second
oscillation     is  completed   the  process     is  repeated  {\it ad
infinitum}. The appearance of periodic pulses, {\it i. e. } scattering
echoes in the decay of the number of trajectories remaining inside the
bottle is evident.

The  period of these echoes can  be  easily extracted by measuring the
intensity of scattered trajectories as they cross some position in the
asymptotic  region.  In  Fig.~\ref{fig:bottle-echoes}-$b$  we plot the
normalized outgoing flux  of scattered trajectories  as a  function of
time  $P(t)$ measured at   $x=20$.  The scattering  echoes are clearly
visible and we obtain   a value for their  period  of $\tau =  7.9 \pm
0.7$. The decay underlying the oscillations in  $P(t)$ is of power-law
nature, with a power of $\approx 2$ which is  well within the expected
limits \cite{power-law-decay}. It is worthwhile mentioning, that we do
not claim that this power law is asymptotic, but it seems well defined
for  almost two decades. We  detect at least twenty oscillations, that
represent     echoes,    so these      correspond  to   a  significant
transient. Diffusion in the chaotic layer spreads the remaining packet
and thus damps the echoes.

As discussed in the previous section, before comparing $\tau$ with the
value predicted by eq.~\ref{eq:magic_2}, we must  divide $\tau$ by the
mean return  time to the surface   of section $\tau_{SOS}$.  Measuring
the return time to the surface of section of the trajectories rotating
around  the   stable   island     we  have    numerically     obtained
$\tau_{SOS}=0.635$ and then, $\tau/\tau_{SOS}=12.44  \pm 1.1$ which is
in  excellent agreement  with   theoretical prediction  of   $T=12.5$.
However, there does not exist a direct procedure by which $\tau_{SOS}$
can be experimentally obtained  from  asymptotic data.  Therefore,  it
would  be desirable  to  test our  method using exclusively quantities
that can be measured or calculated  from asymptotic experimental data.
Unfortunately  this  seems  to  be quite   difficult. Using techniques
developed  in \cite{buetikofer00} we can  determine  the period of the
outer unstable orbit  easily,  but that  is  to crude an estimate.  We
shall have to use some physical understanding of  the system, to get a
typical time  for interior evolution. In the  case of a billiard, this
would be the  transversal  size of the scattering  region,  and we may
take $\tau_{SOS}  \approx  2\Delta$, where $\Delta  $  is  the biggest
width of the bottle.  In this way we obtain a period of $\tau=11.3 \pm
1.0$ which is still  in agreement with  the theoretical expectation if
one recalls that $T$ is also a mean  value with an associated error of
$\pm  0.5$. If   the billiards walls  are  thin  this  number could be
extracted  from a scattering  measurement,   but obviously we have  no
guarantee, that this is true. So we have to have some physical insight
at this point, which  depends very  much on  the system. On  the other
hand  the prediction, that  scattering  echoes are associated with low
developed horseshoe does not depend on this point. We shall see later,
that  in  situations  where wave mechanics   apply we  may  be able to
overcome this problem.

\begin{figure}[!t]
\begin{center}
\includegraphics[scale=0.6]{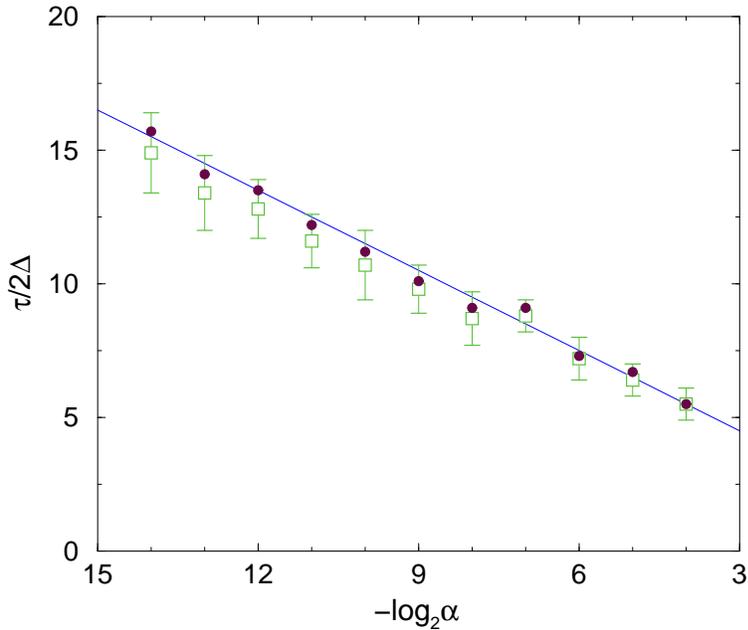}
\caption{\label{fig:bottle-period}  Period  of  the  classical  echoes
 $\tau$  for  the  bottle   billiard  for  the  parameters  listed  in
 Tab.~\ref{tab:bottle}.   We show  both $\tau/2\Delta$  (open squares)
 and $\tau/\tau_{SOS}$ (solid  circles). The straight line corresponds
 to    the    theoretical    expectation    $T(\alpha)$    given    in
 eq.~\ref{eq:magic_2}.}
\end{center}
\end{figure}

To further test the method  of Sec.~\ref{sec:method}, we have measured
the  period  of the  classical   echoes  for  different  sets  of  the
billiard's parameters giving rise to horseshoes at different stages of
development.  Fixing  the values of $\beta  = 1$ and $\gamma=0.85$ and
varying the length   of the stable  periodic  orbit $\Delta$,  we have
obtained horseshoes in stages  of development from  $\alpha=2^{-4}$ to
$\alpha=2^{-14}$.  For each   case  we have measured  the   normalized
outgoing flux  $P(t)$  and from  it  obtained a  value  for the period
$\tau$ of   the   scattering  echoes.  In   table~\ref{tab:bottle}  we
summarize  the different realizations of  the billiard.  We  do so, by
comparing $\tau$ in units  of both, the  period of the stable periodic
orbit $2\Delta$ and the  mean return time   to the surface of  section
$\tau_{SOS}$.  In Fig.~\ref{fig:bottle-period} $\tau$ is compared with
the  theoretical expectations of  eq.~\ref{eq:magic_2}.   We  want  to
remark  that even when  in  all  cases  we  find good  agreement  with
eq.~\ref{eq:magic_2} even better  agreement  is always found  when the
mean $\tau_{SOS}$ is used.

\begin{table}[!t!]
\begin{center}
\begin{tabular}{|c|c|c|c|c|}
\hline
$\Delta$ & $T(\alpha)$ & $\tau/2\Delta$ & $\tau/\tau_{SOS}$ & $\tau_{SOS}$ \\
\hline
\hline
$0.6300$ & $5.5$  &  $5.5 \pm 0.6$ & $5.5  \pm 0.6$ & $1.246$ \\
$0.5925$ & $6.5$  &  $6.4 \pm 0.6$ & $6.7  \pm 0.6$ & $1.135$ \\
$0.5100$ & $7.5$  &  $7.2 \pm 0.8$ & $7.3  \pm 0.8$ & $0.999$ \\
$0.4730$ & $8.5$  &  $8.8 \pm 0.6$ & $9.1  \pm 0.7$ & $0.920$ \\
$0.4550$ & $9.5$  &  $8.7 \pm 1.0$ & $9.1  \pm 1.0$ & $0.862$ \\
$0.4100$ & $10.5$ &  $9.8 \pm 0.9$ & $10.1 \pm 0.9$ & $0.791$ \\
$0.3810$ & $11.5$ & $10.7 \pm 1.3$ & $11.2 \pm 1.4$ & $0.730$ \\
$0.3665$ & $12.5$ & $11.6 \pm 1.0$ & $12.2 \pm 1.0$ & $0.697$ \\
$0.3577$ & $13.5$ & $12.8 \pm 1.1$ & $13.5 \pm 1.1$ & $0.677$ \\
$0.3250$ & $14.5$ & $13.4 \pm 1.4$ & $14.1 \pm 1.4$ & $0.618$ \\
$0.3082$ & $15.5$ & $14.9 \pm 1.5$ & $15.7 \pm 1.5$ & $0.587$ \\
$0.2800$ & $19.5$ & $18.2 \pm 0.9$ & $19.3 \pm 0.9$ & $0.529$ \\
\hline
\end{tabular}
\caption{ Comparison between the values obtained for the period $\tau$
of the scattering  echoes for the  bottle billiard eq.~\ref{eq:bottle}
and those   predicted by eq.~\ref{eq:magic_2} $T(\alpha)$.   We report
$\tau$ in units of the period  of the stable  periodic orbit (given by
$2\Delta$)  and in units  of the  mean  return time to the  surface of
section $\tau_{\rm SOS}$.  We report  the parameter $\Delta$ for which
the  horseshoes  in different  stages  of development ($-\log_2\alpha$
from $4$  to $14$  and $18$)  are obtained.   The obtained  values for
$\tau_{\rm SOS}$  are shown  in last column.   In all cases  the other
parameters were fixed to $\beta = 1$ and $\gamma=0.85$.}
\label {tab:bottle}
\end{center}
\end{table}

For the case of scattering systems giving rise to a ternary horseshoe,
we  shall turn  our attention  in the  next  section mainly  to kicked
one-dimensional  models   described   by the   generic  Hamiltonian of
eq.~\ref{eq:kicked-hamiltonian}.       In     a      previous   letter
Ref.~\cite{mejia01} we presented  a short discussion of the appearance
of classical and quantum scattering echoes in a kicked one-dimensional
system that gives  rise to a  binary horseshoe. The system in question
is described by potential profile given by
\begin{equation} \label{eq:potential-binary}
V(q) = \left\{
\begin{array}{ccl}
\frac{\textstyle q^2}{\textstyle 2} + 1 & , & q < 0 \\
& & \\
e^{-q}(q^2 + q + 1) & , & q \geq 0 \\
\end{array}
\right..
\end{equation}
For the detailed properties of the dynamics of this model we refer the
reader to  \cite{mejia01}.  The technical   details  of the  numerical
computations will be discussed in the next  section, in the context of
a detailed  treatment of ternary   horseshoes.  In order  to gain more
insight  in the origin of  the scattering echoes and more importantly,
on the essential differences between the classical and quantum echoes,
we have followed the evolution of the density distribution function in
phase  space   $\rho_{\rm   cl}(q,p)$   for   the  model   system   of
eq.~\ref{eq:potential-binary}.  Here, $q$ and  $p$ are the canonically
conjugated position and momentum variables.

We have created two animated movies, one for the classical density and
other for an analogous quantum density in terms of the Husimi function
\cite{husimi}.  We first turn our attention to the classical case.  In
movie~1,  we  show  the  time   evolution  of the  classical   density
distribution initially  consisting on a  narrow packet of trajectories
as a stroboscopic map on phase  space.  This distribution is presented
in a  quadratic colour-scale from blue   to red.  For  each frame, the
colour-code is rescaled so that it is possible to follow the evolution
of the ever smaller density that is left inside the scattering region.
Superposed  in a gray-scale,   the corresponding static phase portrait
with the fractal layer of islands and KAM tori is shown.  The value of
$A=0.967$  ({\it c.f.}  eq~\ref{eq:kicked-hamiltonian}),  used in this
simulation was chosen because somewhere  between this value and $A=1$,
the outermost   KAM surface that  appears close  to the outermost dark
structure surrounding the inner fixed point disintegrates.
 
The {\it  movie} starts  at   the time  when the  packet  reaches  the
interaction  region  along the incoming  stable  manifold.  In a first
passage,  part of the packet bounces  on the outside of the hyperbolic
fixed point at $(q,p)=(1,0)$. The rest enters the potential well where
the packet moves around the inner fixed point at $(q,p)=(0,0)$.  After
one rotation  is completed, most  of the packet leaves  the scattering
layer along the outgoing unstable manifold.  A small part stays in the
region of the  homoclinic  tangle  and continues rotating   around the
island.   We then see the density  concentrated in a narrow region and
slowly diffusing into the fractal zone of islands around the outermost
KAM surface.   This very diffusion   also kills the echoes  relatively
rapidly  because  it widens  the  trapped  part  of the initial packet
systematically.  It is also  responsible for the power-law decay, that
is characteristic of  such systems.  A  careful observation shows that
each time the  packet  reaches the region  around  the unstable  fixed
point, part of the density is scattered out forming  an echo while the
rest continues rotating with the scattering layer.

The process  of emission  of  the echoes  appears more clearly  in the
evolution of   the quantum density.  Under  conditions  similar to the
classical case, we show  in movie~2, the time  evolution of the Husimi
distribution of a Gaussian wave packet for $A=0.967$, $\hbar=0.01$ and
$\sigma=2.5$,  initially at  $q_0=100$   and $p_0=-1.48$  ({\it  c.f.}
eq.~\ref{eq:wave-packet}).  The Husimi distribution  is presented in a
quadratic colour-scale from blue to  red, and again the colour-code is
rescaled for each frame.

The direct  part of the  scattering of the  distribution evolves quite
similar to the  classical case, bouncing off or  going once around the
island.   The trapped part  on the other hand  shows a quite different
behaviour.   It  penetrates  the  fractal  area   rapidly, so  we  can
conclude,   that tunneling  rather  than   diffusion is  the  dominant
mechanism, and  indeed  this  process   quickly extends  inside    the
classically forbidden region of the central stable island. As we shall
corroborate in next section, in the quantum case the echoes detect the
winding numbers inside the outermost KAM-surface instead of those that
characterize  the scattering  layer.  For our  examples  and indeed in
typical situations these are larger than on the outside, and therefore
we can expect  to see  shorter periods as  we have  the opportunity to
look {\it  inside} the island.   However, by  decreasing the  value of
$\hbar$,  the period of the  quantum echoes will approach the estimate
of  eq.~\ref{eq:magic_2} leading to  the possibility of a quantitative
estimate of the development of the classical horseshoe.  In movie~2 we
clearly see how part of the quantum probability  density is emitted in
echoes.   We  also see that  the formerly   narrow packet that tunnels
inside the island spreads in time over the  invariant surface where it
is trapped.  This has the effect  of gradually increasing the width of
the echoes,  though they persist   much longer, than in the  classical
case.  Again the observation of the scattering echoes will be possible
only in an intermediate, though possibly quite long, time scale.

\subsection{Ternary horseshoe}
\label{sec:ternary-horseshoe}

We now turn our attention to ternary horseshoes.  We shall concentrate
on  a second   type of  models   that  is  on  kicked  one-dimensional
scattering systems.   The advantage of such  models consists mainly in
the fact that  the dynamics is easily solved  both in classical and in
wave mechanics.

The model we study is described by  a time dependent Hamiltonian given
by
\begin{equation} \label{eq:kicked-hamiltonian}
H(q,p,t) = \frac{p^2}{2}  + A \  V(q) \sum_{n=-\infty}^{\infty} \delta
(t - n) \ .
\end{equation}
The time dependence  consists of an  infinite periodic train  of delta
pulses   kicking with  period    $1$  the  otherwise  free  particle's
trajectory.  The physical parameter $A$ determines the strength of the
kick.  Such  periodically driven  one-dimensional systems are formally
equivalent to time-independent Hamiltonian systems with two degrees of
freedom \cite{lichtenberg91}.  We thus expect to test the more general
situation with reduced effort.

As Poincar\'e map for the classical problem we choose the stroboscopic
map at times $t = n + 1/2$
\begin{equation} \label {eq:stroboscopic-map}
\begin{array}{lcl}
p_{n+1} & =  & p_n - AV'(q_n+p_n/2) \\  q_{n+1} & = & q_n +  p_n - A
V'(q_n+p_n/2)/2 \ .
\end{array}
\end{equation}
where $V'$  stands for the derivative  of the potential term. It gives
the evolution of  a  classical trajectory  from time $n-1/2$  to  time
$n+1/2$ and is symmetric with respect to time inversion.

\begin{figure}[!t]
\begin{center}
\includegraphics[scale=0.6]{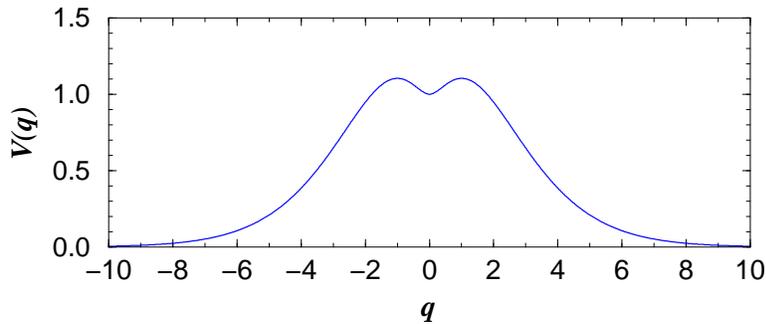}
\caption{
\label{fig:k-t-potential}
Potential profile of eq.~\ref{eq:k-t-potential}.}
\end{center}
\end{figure}

The model in question is determined by a potential profile given by
\begin{equation} \label{eq:k-t-potential}
V(q) = e^{-|q|}(q^2 + |q| + 1)
\end{equation}
It consist on  a finite potential well  centered at $q=0$  enclosed by
two potential barriers at $q=\pm 1$. At infinity, the potential decays
exponentially   to     zero.     This    potential is     shown     in
Fig.~\ref{fig:k-t-potential}.     The  system  has three   fundamental
periodic  orbits and therefore,   gives rise  to a  ternary horseshoe.
Again, the  stability of the  interior periodic orbit at $q=0$ depends
on the strength of the  potential  $A$.  The exterior periodic  orbits
corresponding to the potential barriers, are unstable for any value of
the physical parameter. In a phase  portrait they appear as hyperbolic
fixed points at $(q,p)=(\pm 1,0)$.  The ternary horseshoe is traced by
the stable and unstable manifolds of both hyperbolic points.

\begin{figure}[!t]
\begin{center}
\includegraphics[scale=0.75]{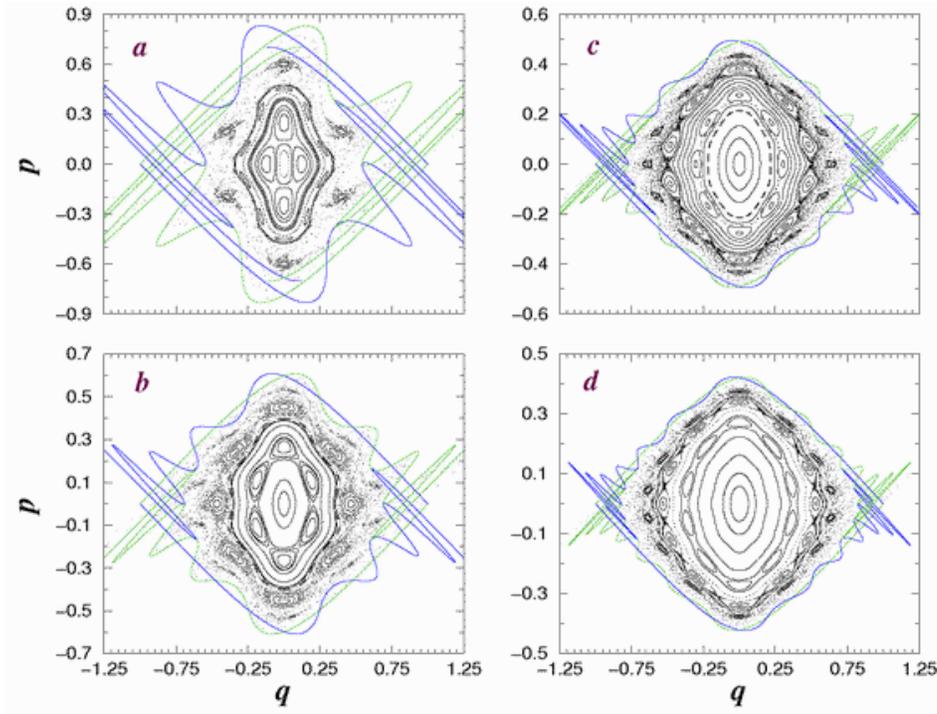}
\caption{\label{fig:t-k-phase-portrait}  Horseshoe and  phase portrait
for the model system eq.~\ref{eq:k-t-potential}, for several values of
the formal development parameter  $-\log_3 \beta$: a) $3$ for $A=2.6$,
b) $5$ for $A=1.54$, c) $7$ for $A=1.065$ and d) $9$ for $A=0.8$.}
\end{center}
\end{figure}

We have extensively analysed the scattering process of this system for
several values of the physical  parameter $A$ and thus, for horseshoes
in     different   stages        of        development.             In
Fig.~\ref{fig:t-k-phase-portrait}, we show the  phase portrait and the
underlying ternary horseshoe for four different values of $A$.  We can
observe how the horseshoe development decreases from $\beta=3^{-3}$ to
$\beta=3^{-9}$ with  the value of  $A$.  This  occurs together with an
enlargement of the stable island. The more  developed the horseshoe is
the    more extended the  scattering  layer.  These  cases fulfill the
conditions under  which the appearance  of   echoes is expected,  {\it
i.e.}, in all cases  of Fig.~\ref{fig:t-k-phase-portrait}, there exist
a large  stable central island surrounded by  a well developed chaotic
scattering layer.   Note also the decrease of  the winding number with
$A$ in  terms of the  large chain of  secondary islands living  at the
interior  of the island. The stable  island disappears at  $A=4$ as it
becomes unstable.  Then  the horseshoe is  close  to a development  of
$\beta=1/3$  and reaches  complete development  and thus hyperbolicity
near $A=14.3225$.

\begin{figure}[!t]
\begin{center}
\includegraphics[scale=0.6]{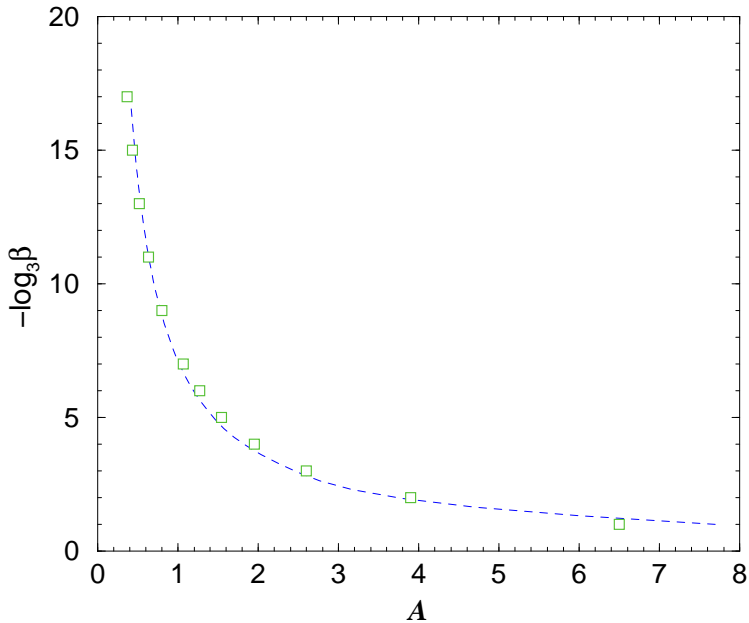}
\caption{
\label{fig:k-t-development}
Development  of the horseshoe  $-\log_3(\beta)$ as  a function  of the
strength    of   the    potential   $A$    for   the    potential   of
eq.~\ref{eq:k-t-potential}.   For the  considered values  of  $A$, the
squares  correspond  to  the   development  $\beta$  of  the  obtained
horseshoe.    The  line  corresponds   to  $\beta(A)$   obtained  from
eq.~\ref{eq:magic_3},  from the  numerically  determined mean  orbital
period $T(A)$.}
\end{center}
\end{figure}

Before  discussing the  determination of the period of the  scattering
echoes we have  numerically computed the mean orbital  period $T$ as a
function  of  the  physical  parameter  $A$.  To  this  end,  we  have
considered  trajectories targeting  the potential  well  from outside.
This is important  as we want to probe  exclusively winding numbers of
the    scattering    layer.     We    recall   that    by    inverting
eq.~\ref{eq:inv_magic_3}, we obtain an  expectation value for the mean
orbital period of the  scattering trajectories rotating in the chaotic
scattering layer close to the surface  of the island. As a function of
the formal development parameter $\beta$, the orbital period
\begin{equation} \label{eq:magic_3}
T = -2\log_3 \beta + \frac{3}{2} \ , \quad \quad \beta = 3^{-n}
\end{equation}
is, as before, given in units of $\tau_{SOS}$.  Plugging in the values
calculated for     $T(A)$  into eq.~\ref{eq:magic_3}  we    obtain the
development parameter $\beta$ as a  function of the physical parameter
$A$. In Fig.~\ref{fig:k-t-development} we  plot $\beta(A)$ as a dashed
line.  The behaviour of $\beta(A)$ coincides  with the observations of
Fig.~\ref{fig:t-k-phase-portrait} in which the horseshoe develops with
$A$.  Furthermore, we have directly  determined the development of the
horseshoe  for several  values of $A$.  These values  are shown as the
squares  in Fig.~\ref{fig:k-t-development}.    The  agreement with the
behaviour  of $\beta(A)$ determined from   the orbital period supports
the identification of  the period of the  echoes with the mean orbital
period.

Note that  this result can be rather  simply generalized, for the case
of   an  asymmetric ternary   horseshoe with two   outer  fixed points
yielding
\begin{equation} \label{eq:magic_3as}
T = -\log_3 \beta _1 -\log_3 \beta _2
 + \frac{3}{2} \ .
\end{equation}
To invert this equation we will need the relative  phase of the echoes
at both openings of the billiard;  this can be achieved experimentally
for open billiards, while it might be more difficult in other systems.

\begin{figure}[!t]
\begin{center}
\includegraphics[scale=0.45]{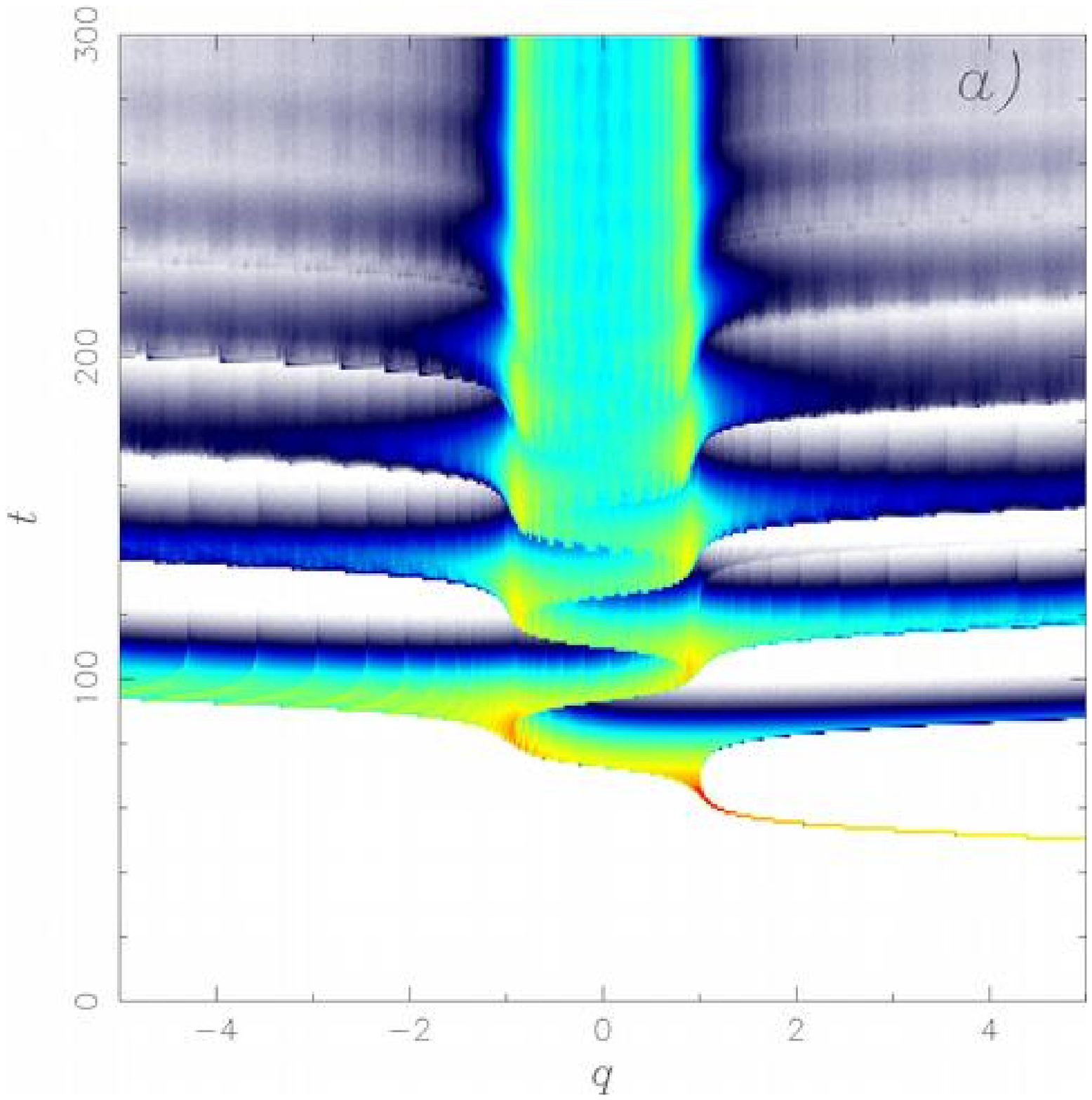}
\includegraphics[scale=0.4]{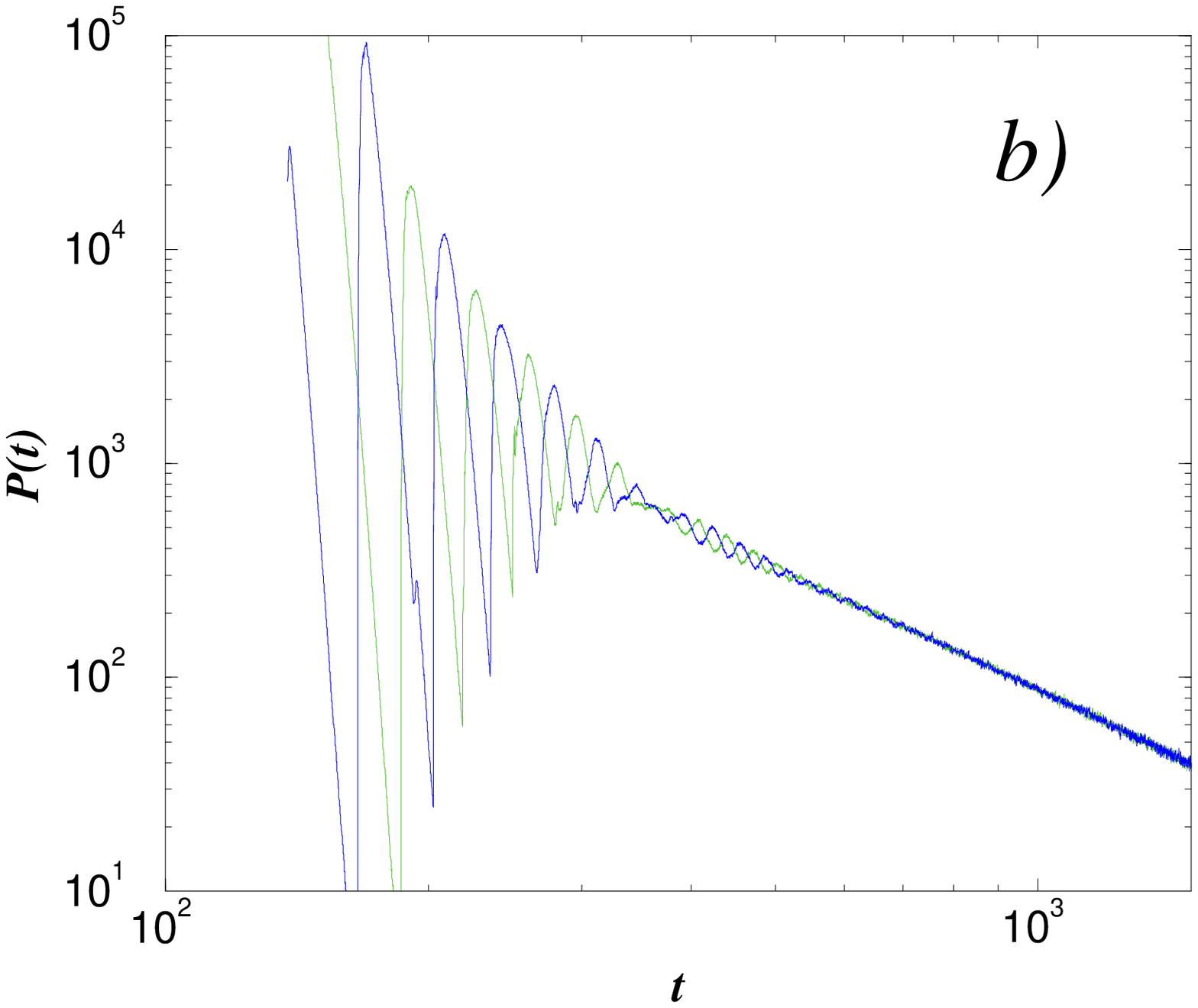}
\caption{\label{fig:k-t-classical-echoes}   $a$)   Classical   density
distribution    in $\rho_{\rm   cl}(q,t)$    for   the  model   system
eq.~\ref{eq:k-t-potential})  in  a   logarithmic   colour scale    for
$A=0.366$  corresponding  to a development of    $-\log_3 \beta = 17$.
$b$)  Outgoing flux of classical   trajectories as a  function of time
$P(t)$, measured at $q=-20$ (green) and $q=20$ (blue).}
\end{center}
\end{figure}

We shall now look at simulations to study the scattering echoes of the
proposed symmetric ternary horseshoe. The  situation is quite  similar
to the one discussed above for a binary horseshoe, except that the two
external fixed points imply  that the trajectories rotating around the
stable island can now leave the scattering layer to both sides.  Again
we have prepared a narrow packet of  initial conditions in phase space
placed   at the  asymptotic region.    Solving  the dynamics   through
eq.~\ref{eq:stroboscopic-map} we follow the evolution of the classical
density distribution $\rho_{\rm  cl}(q,t)$ in the  $q$-$t$ plane.   In
Fig.~\ref{fig:k-t-classical-echoes}-$a$, we show  $\rho_{\rm cl}(q,t)$
for $A=0.366$. The  packet was initially centered at $(q,p)=(20,0.9)$,
deep in the asymptotic region.

As in the binary case,  the part of the  classical density that enters
the   potential well  oscillates  inside and   decays  in echoes.  The
description of  the scattering process is exactly  the same as  in the
case of  the binary horseshoe except  that now the echoes  are emitted
with a period of $T/2$ to one side and  the other. As a consequence of
the way we  target the potential, if  we measure the  outgoing flux of
trajectories $P(t)$ at symmetric positions  then we shall observed the
echoes  in      counter    phase.    This   can     be    observed  in
Fig.~\ref{fig:k-t-classical-echoes}-$b$     in  which  we  plot $P(t)$
measured at $q=\pm20$. The decay underlying the oscillations in $P(t)$
is again of power-law nature, with a power of $\approx 2$.

Before testing  our theoretical  expectation  for  the  period of  the
echoes we consider   the  quantum scattering process  for  this kicked
system. For  this purpose we  use the  unitary time evolution operator
easily obtained  for kicked systems.   For  one time  step it  can  be
written    as  three  phases    intertwined   by  Fourier   transforms
$\mathcal{F}$, which take  us from coordinate   to momentum space  and
back.   Thus  we   obtain  for   the  kernel  of   this   operator  in
momentum-space

\begin{equation}
\label {u-time-evolution} U (p',p) = \exp\Big[{-\frac{i}{4\hbar} \,
{p^2}}\Big] \ \mathcal{F} \ \exp \Big[\frac{i}{\hbar} \, AV(q)\Big] \
\mathcal{F}^{-1} \, \exp\Big[{-\frac{i}{4\hbar} \, {p^2}}\Big] \ .
\end{equation}
This expression  is simple and very  efficient if good fast
Fourier transform (FFT) codes are used.

We   shall analyze  our  scattering  system  in terms  of  wave packet
dynamics, and use minimum uncertainty Gaussian wave packets given by
\begin{equation} \label {eq:wave-packet}
\Psi(q,0) = \frac{1}{\pi^{1/4}\sigma^{1/2}}
\exp \Big[{-\frac{(q-q_0)^2}{2\sigma^2} +
\frac{i}{\hbar}p_0 q}\Big] \ .
\end{equation}
For   a given  value  of the  initial   momentum $p_0$, $\sigma$  will
determine  the  duration of  the  pulse. The   value of  $\hbar$  will
determine how   near to the  classical limit  we operate.  Recall that
short pulses imply a poor energy resolution,  while long ones can have
fairly well  defined energies.  We can therefore  use short pulses and
consider the time evolution, or we can use long pulses and look at the
energy dependence of some outgoing quantity.  We  shall start with the
former.

\begin{figure}[!t]
\begin{center}
\includegraphics[scale=0.45]{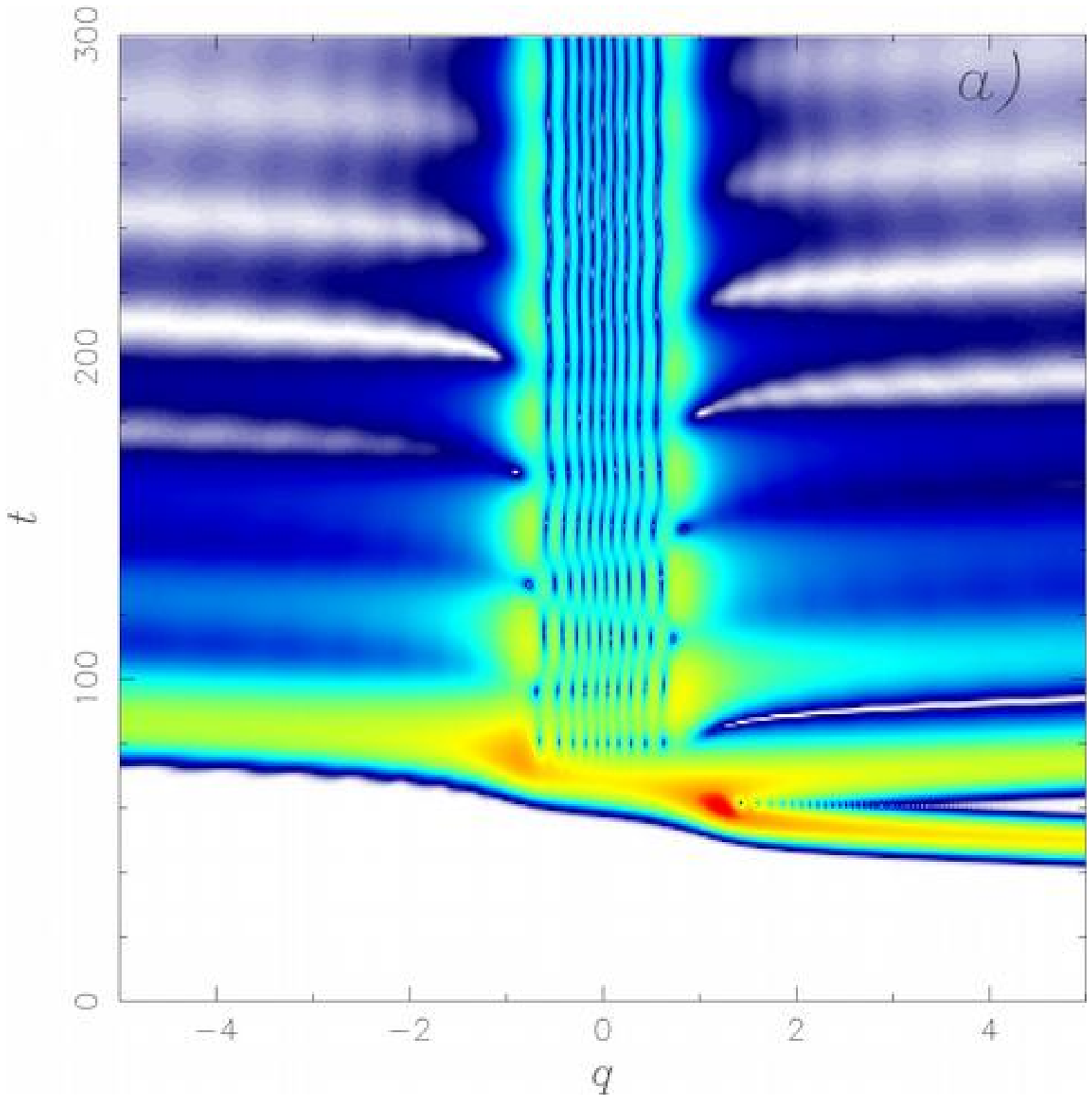}
\includegraphics[scale=0.4]{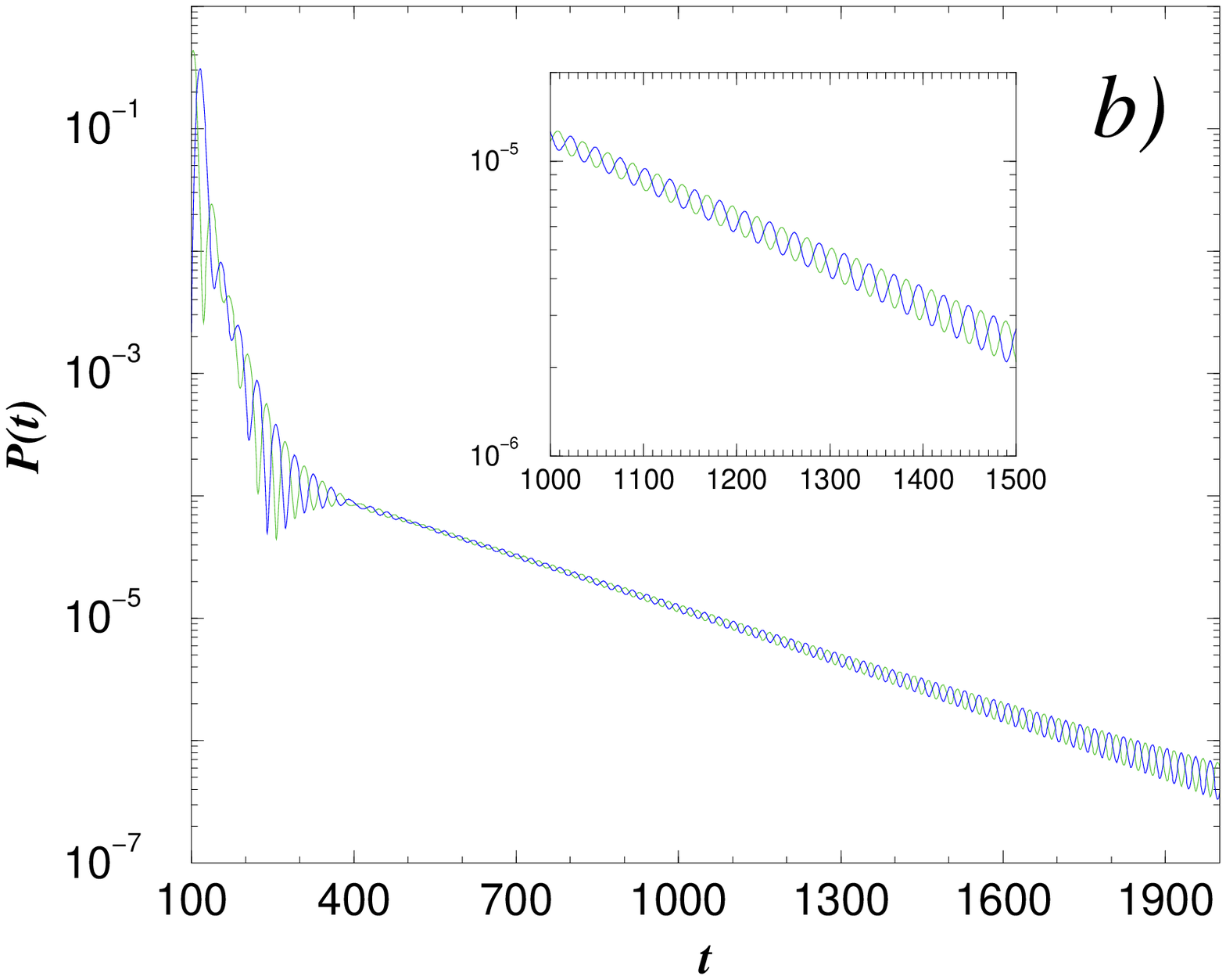}
\caption{\label{fig:k-t-quantum-echoes}   {\it  a})   Quantum  density
distribution function in  configuration space $\rho_{\rm q}(q,t)$ as a
function of time for the model system of eq.~\ref{eq:k-t-potential} in
a   logarithmic    colour  scale for    the   same  parameters   as in
Fig.~\ref{fig:k-t-classical-echoes}.    The incident  wave-packet  was
initially set at $q=50$ and  energy $E_{in}=0.40415$ corresponding  to
the energy domain of the incoming manifold of the classical horseshoe.
$\hbar=0.01$ and $\sigma=2.5$.  {\it b}) Outgoing flux $P(t)$ measured
at $q=-20$ (green) and $q=20$ (blue).}
\end{center}
\end{figure}

In  close     analogy   to    the  classical    case   we   show    in
Fig.~\ref{fig:k-t-quantum-echoes}-$a$    the    quantum    probability
distribution in configuration space  $\rho_{\rm q}(q,t)$ as a function
of     time in a     logarithmic   colour-scale  code  for  $A=0.366$,
$\hbar=0.01$.  The wave  packet with $\sigma=2.5$ was initially placed
at  $q_0=50$ with an  incoming energy $E_{\rm in}  = 0.40415$ Again we
clearly distinguish the incoming pulse, the part directly scattered at
the barrier  and another part that entered  the well of the potential,
but  leaves   directly to the    left.   At longer    times we see  an
oscillating  packet  inside the well  mimicking the  behaviour  of the
classical density distribution but with  an amplitude that corresponds
to the classically  forbidden region in  phase  space.  Each  time the
wave packet returns to either one of the borders of the potential well
({\it i.e} $q=\pm 1$) an ``echo'' is emitted to infinity.  This occurs
alternately to the left and right of  the potential well with a period
that corresponds  to  half  the  orbital  period.  Since  the  quantum
probability   is  rotating inside the  stable   island the information
encoded  in  the  period  of  the   quantum  scattering echoes    will
corresponds  to a winding number characteristic  of the  region of the
island on which the   probability is rotating.  Therefore, we   expect
that due to quantum tunneling the period of the quantum echoes will be
smaller than that of the classical case.

Again, in order   to obtain a   value  for the period of   the quantum
scattering echoes, we measure the outgoing flux as  a function of time
$P(t)$,  measured     around   $q=\pm    20$.  This   is     shown  in
Fig.~\ref{fig:k-t-quantum-echoes}-$b$.  The echoes emitted to left and
right are  in  counter phase as  expected.   The decay of  the quantum
probability is exponential.  This occurs   at values of $A$ for  which
the  classical decay is power-law as  expected for a mixed phase space
situation and is an additional signature  that tunneling has occurred.
It  is  worthwhile mentioning that  in  our calculations   we have not
observed the $1/t$ power law  regime  for intermediate times as  found
and discussed in \cite{casati99}.

\begin{figure}[!t]
\begin{center}
\includegraphics[scale=0.6]{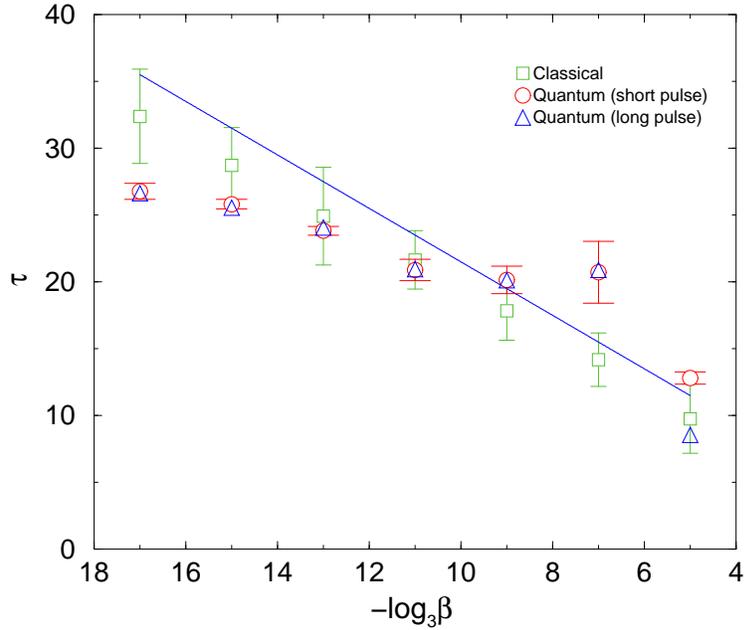}
\caption{Period of  the classical and  quantum echoes $\tau$,  for the
model system of  eq.~\ref{eq:k-t-potential}. The squares correspond to
the period of  the classical echoes. The period  of the quantum echoes
was calculated  from scattering  simulations with short  (circles) and
long (triangles)  pulses.  The straight line corresponds  to $T$ given
in eq.~\ref{eq:magic_3}.
\label{fig:k-t-period}
}
\end{center}
\end{figure}

We show quantum simulations with short pulses $\sigma=2.5$ for several
values of the strength parameter $A$.  In Fig.~\ref{fig:k-t-period} we
summarize  our findings comparing  the obtained values for the quantum
echoes with the classical  echoes and with the theoretical expectation
of eq.~\ref{eq:magic_3}.   We   display the  obtained  period   of the
classical (squares) and quantum (circles) echoes  as a function of the
development parameter of  the ternary horseshoe $\beta$  directly read
off the  horseshoe for the corresponding  values of the parameter $A$.
At low developments stages of the  horseshoe the period of the quantum
echoes   is  consistently shorter than  for  the  classical echoes. As
discussed, this shortening    is   to be  expected  as   the   quantum
probability  can tunnel through  the  classical invariant KAM surfaces
and thus, test  orbital periods inside  the island.  For the model  in
question the typical situation  holds, in which this  period decreases
with the  distance to   the  center of the    island. This  result  is
interesting by itself as the  quantum scattering echoes could be  used
to scan the dynamics in the interior of the stable  island which is by
no means,  accessible to classical experiments.  We shall discuss this
in more  detail  below.   For cases  in  which the  horseshoe  is more
developed the period  of the quantum echoes  turned to  be longer than
for the classical  echoes.  This may be a  particular artifact  of the
structure of the chaotic layer of this model  but, in any case here we
approach  to the region  in  which our  theory ends to  be valid.   In
Fig.~\ref{fig:k-t-period} the  straight    line  corresponds  to   the
expected   behaviour of    eq.~\ref{eq:magic_3}.  The  period   of the
classical  echoes is  found to be  in  good agreement within numerical
accuracy.

\begin{figure}[!t]
\begin{center}
\includegraphics[scale=0.6]{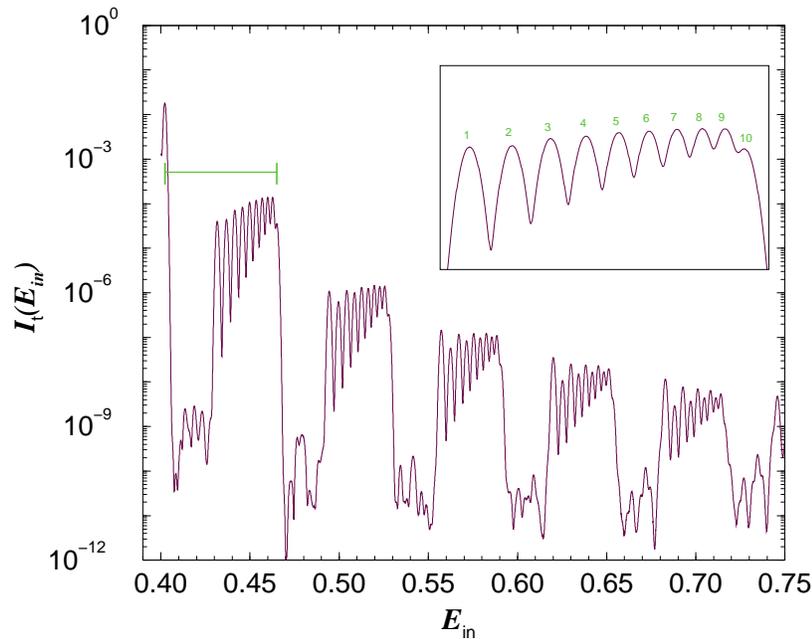}
\caption{\label{fig:k-t-long-pulses}   Quantum  probability  amplitude
$I_t$ inside  the potential well  as a function  of the energy  of the
incident wave packet $E_{in}$ measured  at time $t=500$ for long pulse
with   $\sigma=10$.   The   rest   of  the   parameters   are  as   in
Fig.~\ref{fig:k-t-quantum-echoes}.   The  green  bar correspond  to  a
width  of $2\pi\hbar$  from the  resonance  that occur  at the  energy
domain of  the incoming  manifold of the  classical horseshoe.  In the
Inset  an amplification  of the  first group  of resonances  have been
labeled for discussion purposes.}
\end{center}
\end{figure}

So far, the incoming wave packets were chosen to be narrow in position
space.  This implies the good resolution obtained for observables as a
function of time. However, we can also use long pulses that are narrow
packets  in momentum space, to obtain  good energy resolution. In this
case we  have chosen as  observable the quantum  probability amplitude
integrated over the potential well region at time $t$ as a function of
the  asymptotic  energy  of  the incident wave   packet $I_t(E_{in})$.
While this quantity shows a resonance spectrum very similar to the the
$S$-matrix  it turned  out in our  numerics, that  we  obtained better
resolution  with this  quantity.  In Fig.~\ref{fig:k-t-long-pulses} we
show $I_t(E_{in})$ measured at $t=500$  with $\sigma=10$ for $A=0.366$
and $\hbar=0.01$.  The time at which  $I_t(E_{in})$ is measured is not
relevant as   long as the    measure is  performed   after the  direct
scattering  has subsided.   If this is   the case,  the change  in the
position of the resonances is negligible  as its amplitude decays very
slowly.  We find a group of narrow resonances periodically repeated in
energy.  The  presence   of these sharp  resonances  is  an additional
indication that tunneling between  the regular and chaotic  regions of
the classical phase portrait occurs \cite{seba93}.  From $I_t(E_{in})$
it is  possible to  extract two energy  scales.  The first and trivial
scale corresponds to the width in energy of the group of resonances or
to be more  precise, to the energy  length at which each  resonance is
repeated.  This energy scale equals  $2\pi\hbar$ and correspond to the
period     of   the  kick      that    we have      taken   $1$.    In
Fig.~\ref{fig:k-t-long-pulses} the  green horizontal bar measures this
energy scale.

\begin{figure}[!t]
\begin{center}
\includegraphics[scale=0.5]{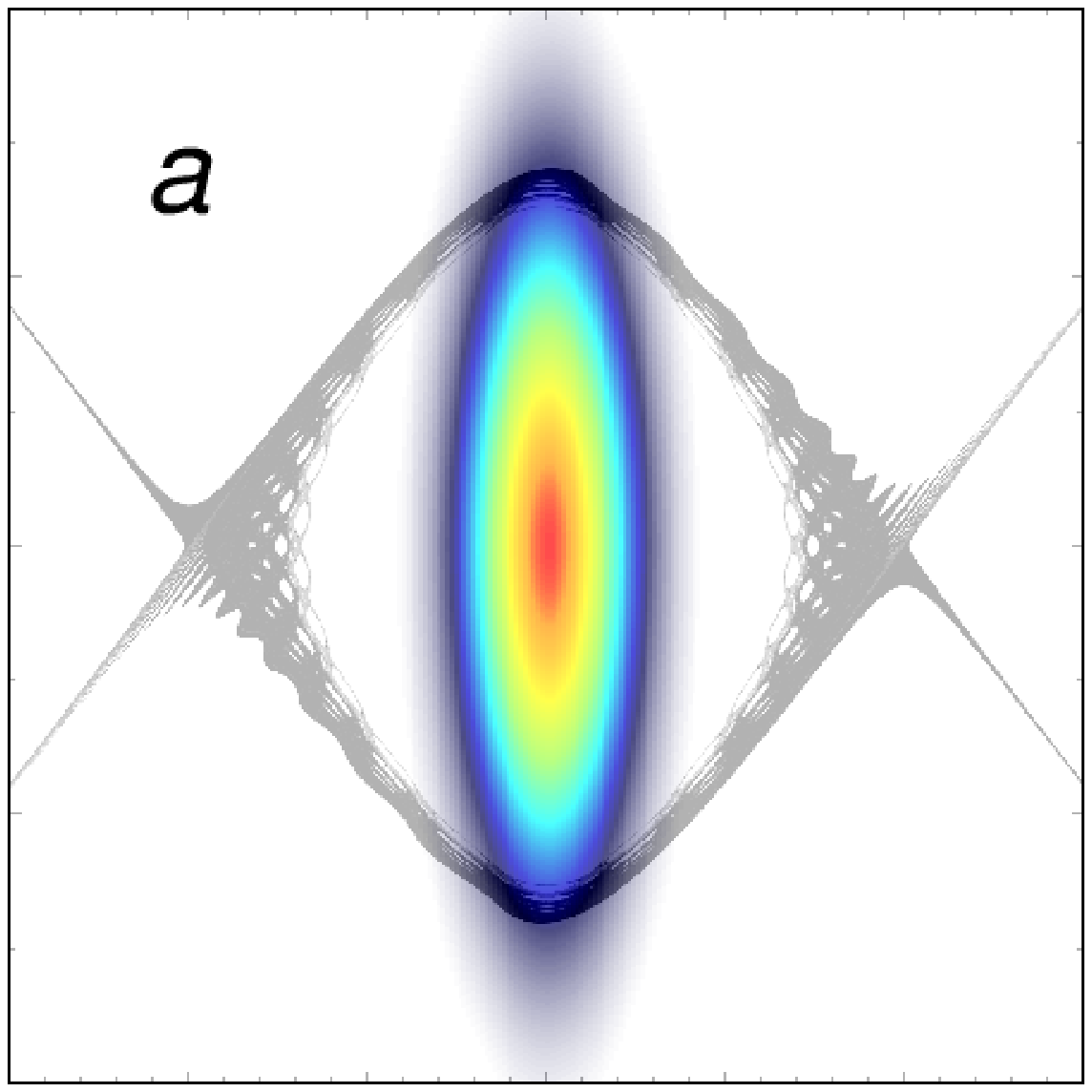}
\includegraphics[scale=0.5]{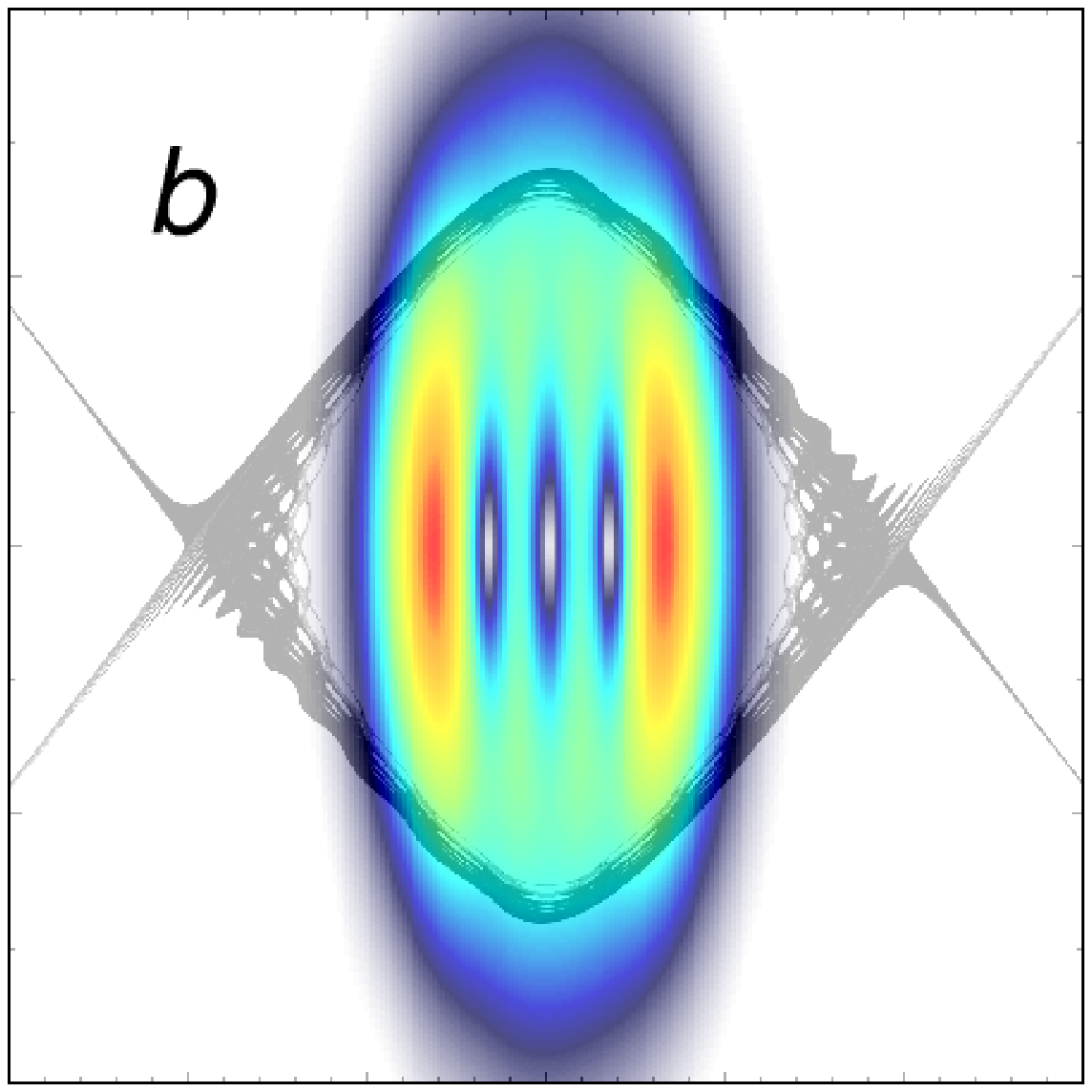}
\includegraphics[scale=0.5]{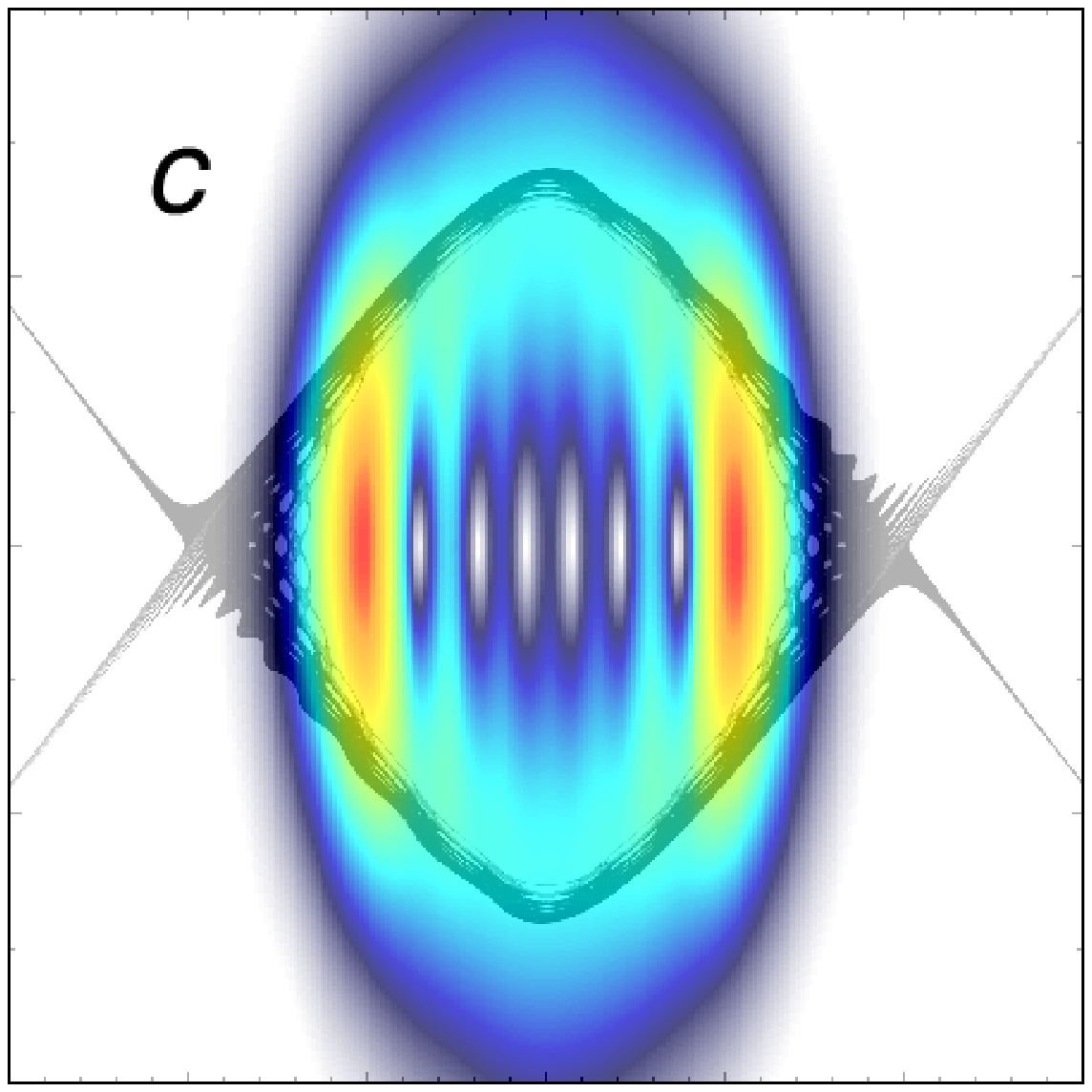}
\includegraphics[scale=0.5]{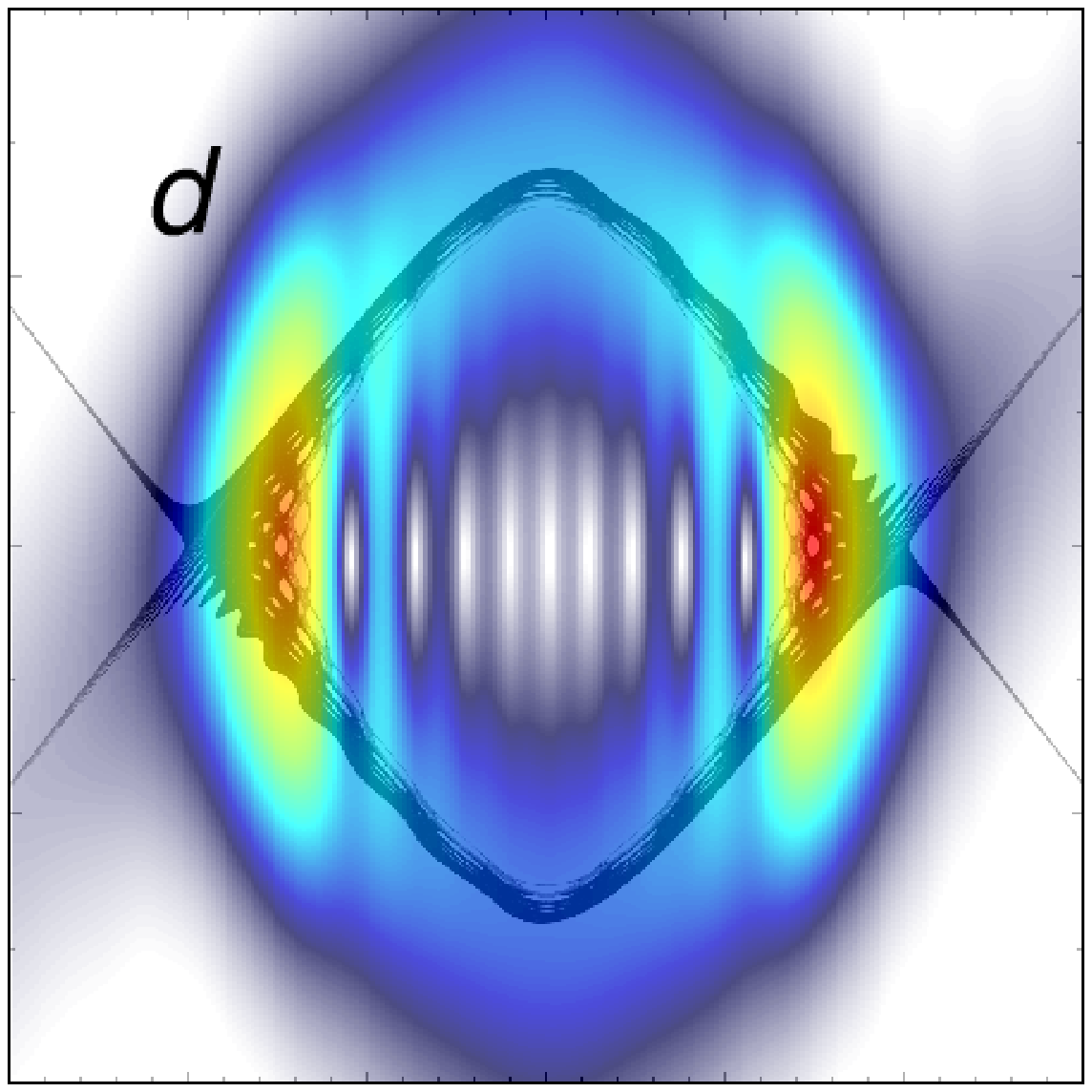}
\caption{\label{fig:k-t-quasibound-states} Husimi distribution for and
incident long  pulse with $\sigma=10.0$ superimposed  to the classical
phase   portrait.    The   rest   of   the  parameters   are   as   in
Fig.~\ref{fig:k-t-quantum-echoes}.  The  energy of the  incident pulse
was chosen as the  energy of resonance: a) $1$, b) $4$,  c) $7$ and d)
$10$.}
\end{center}
\end{figure}

The second  energy scale   that can  be  extracted  from $I_t(E_{in})$
corresponds to the separation between  resonances and, in analogy,  it
necessarily corresponds to $2\pi\hbar/\tau$ with  $\tau$ the period of
the quantum echoes.  However, the energy separation between resonances
decreases with energy.  To  understand this energy dependence  we have
carefully analyzed    each  resonance separately.    In  the  inset of
Fig.~\ref{fig:k-t-long-pulses} an amplification  of the first group of
resonances  is shown.   Floquet  formalism  \cite{floquet}  applies to
systems  that,  like our kicked model eq.~\ref{eq:kicked-hamiltonian},
possess a time periodic potential.  The translation invariance results
in a description in all respects analogous to  the Bloch formalism for
spatially periodic systems.    In the Floquet formalism  these  narrow
resonances can be interpreted as  the quasi-bound states associated to
the system's potential.    To analyse  the  structure the   quasibound
states  have   in phase   space,  we   have calculated  their   Husimi
distribution \cite{husimi}.  We prepare a long  pulse with an incoming
energy equal to the one  of the resonance we want  to study.  At  some
time,  after which the most of  the probability has decayed we compute
the  Husimi  distribution  of   the  quantum  probability  inside  the
potential well.   In Fig.~\ref{fig:k-t-quasibound-states} we show in a
colour scale the obtained Husimi distribution for resonances $1$, $4$,
$7$    and   $10$      as       labeled   in   the       inset      of
Fig.~\ref{fig:k-t-long-pulses}.                                     In
Fig.~\ref{fig:k-t-quasibound-states}  the  Husimis are superimposed to
the corresponding classical phase portrait  in gray.  From this figure
a clear  answer to the energy dependence   emerges.  Each resonance in
the group corresponds to a quasibound state.  On  phase space they are
ring-shape  states  around  the  center  of the  stable  island.   The
quasibound  state  corresponding to  the   first resonance  lives deep
inside the island  while for  the last  resonance $10$ the  quasibound
state  covers the classical scattering  layer.  The more energetic the
resonance is the further out the region on which it lives.  Therefore,
we can conclude that the energy scale associated to  the period of the
scattering  echoes must  correspond to  the  energy separation between
those  resonances associated to  the  most exterior quasibound states,
{\it i.e.}, to  the separation between the last  two resonances in the
group.

Following this strategy we have   obtained $I_t(E_{in})$ for the  same
values  of the parameter $A$  for which  we  have analysed the quantum
scattering in  terms of short pulses   and measured the period  of the
echoes  as $\tau=2\pi\hbar/\Delta  E$  with $\Delta  E$ the separation
between the last two resonances in the group. The results are shown in
Fig.~\ref{fig:k-t-period} as triangles. The correspondence between the
results extracted  from  short  and  long pulses  is  excellent,  thus
opening the possibility to obtain the period of the quantum scattering
echoes in the time or energy domain.  It is worthwhile to mention that
the first large resonance appearing  in the spectrum of  $I_t(E_{in})$
has a classical interpretation.  Indeed  its energy corresponds to the
asymptotic energy of the fringe in which the tendrils emerge along the
outer branches of the  invariant manifolds of the classical horseshoe.
As      indicated   by       the    horizontal   green      bar     in
Fig.~\ref{fig:k-t-long-pulses},  the  position of  this  ``classical''
resonance  is a distance of $2\pi\hbar$  from the last resonance, thus
confirming the association of the period of  the quantum echoes to the
energy separation between the last resonances.

\begin{figure}[!t]
\begin{center}
\includegraphics[scale=0.6]{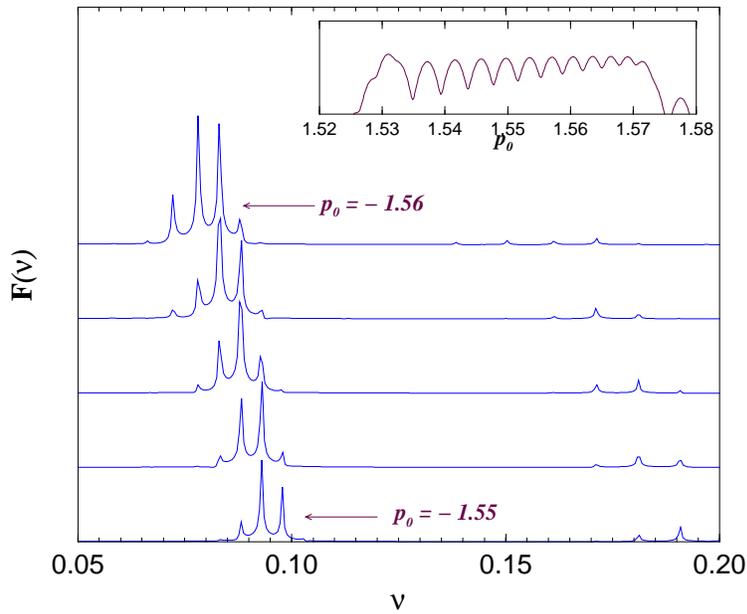}
\caption{\label{fig:beating} Fourier transform of the quantum outgoing
flux  $P(t)$ for  the model  system of  eq.~\ref{eq:k-t-potential} for
$A=1.065$.   The  spectra  correspond  to initial  wave  packets  with
different   incoming  momenta,  from   bottom  to   top:  $p_0=-1.55$,
$p_0=-1.5525$, $p_0=-1.555$, $p_0=-1.5575$  and $p_0=-1.56$.  The rest
of the parameters are  $q_0=50$, $\sigma=2.5$ and $\hbar=0.01$. In the
inset  we plot  the  first  group of  resonances  appearing in  $I_t$,
calculated   at  $t=125$.   The  first   resonance   at  $p_0\sim1.53$
corresponds to the classical resonance.}
\end{center}
\end{figure}

We want   to  call attention  to  the  possibility  to  apply  quantum
scattering echoes as a tool to explore the dynamics in the interior of
the island inaccessible to  classical scattering trajectories. For the
model    system  eq.~\ref{eq:k-t-potential}   we  have performed   the
following  numerical  experiment: As before,  we  target the potential
well with a short pulse $\sigma = 2.5$  with initial incoming momentum
$p_0$ and measure  the outgoing quantum flux  $P(t)$ as  a function of
time.    Applying a Fourier   transformation to $P(t)$   we obtain its
frequency spectrum  ${\mathrm F}(\nu)$.  As  $P(t)$ corresponds to the
decay of the quantum probability that stays rotating around the stable
island its Fourier spectrum simply tell  us the frequency at which the
quantum  probability  rotates.    In  Fig.~\ref{fig:beating}  we  show
${\mathrm F}(\nu)$  for   five initial  wave packets   with  different
incoming momenta.  What we observe is  that the probability trapped in
the potential    well, rotates with a   period  which depends   on the
incoming momentum of  the packet.   Comparing  the five spectra it  is
clear that the  excited    frequencies  correspond to  those  of   the
quasibound states of the system.   Indeed this is the   case as it  is
confirmed in the inset of the same figure where the resonance spectrum
of  $I_t(p_0)$  as a  function  of   the  incoming momentum  is shown.
Therefore, by measuring the period of the quantum echoes obtained from
an incident packet that scans the incoming energy we can produce a map
of  the characteristic  winding  numbers   in   the interior   of  the
island. In a sense, wave mechanics may be used as an "x-ray device" to
study  regions inaccessible  in   classical mechanics.  Under  certain
circumstances this may be a  way to solve the  problem of finding  the
return time to the classical surface of  section we encountered above,
because the differential rotation time will  converge to the period of
the inner periodic orbit, which has the value $2 \Delta$ which we used
successfully   to  approximate this time.   Similar   results could be
obtained  by Fourier transform of  results  in the  energy domain,  as
touched upon in the conclusions.

\begin{figure}[!t]
\begin{center}
\includegraphics[scale=0.6]{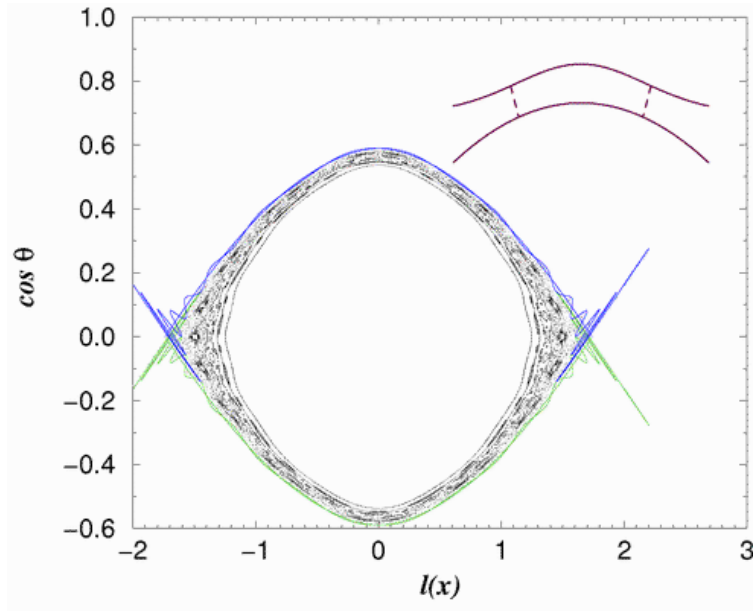}
\caption{\label{fig:knee-horseshoe}  Classical   horseshoe  and  phase
portrait for  the knee  billiard for $\alpha=0.161$,  $\gamma=0.2$ and
$\delta=0.1$.  The  development of the  horseshoe is $-\log_3\beta=8$.
The shape  of the  billiard is shown  at the  top right of  the figure
where the dashed lines correspond to the unstable periodic orbits.}
\end{center}
\end{figure}

As  a final example we briefly  comment the classical  scattering in a
billiard  giving rise  to a  symmetric ternary  horseshoe.  A study of
this particular example and  its experimental realization  will appear
elsewhere  \cite{exp04}.  The billiard in     question consist of  two
boundaries: A Gaussian and a parabola given by
\begin{equation} \label{eq:billiard}
\begin{array}{rcl}
y(x) & = & \exp{(-\alpha x^2)} \\
\\
y(x) & = & \gamma - \delta x^2  \ .
\end{array}
\end{equation}
Both boundaries extend from $-\infty$ to $\infty$.   The shape of this
billiard is shown inside Fig.~\ref{fig:knee-horseshoe} for some set of
its parameters.  This billiard  has three fundamental periodic orbits,
one in  each neck and one  in the center.   The stability of the inner
periodic orbit   corresponding to  a  vertical  trajectory  at  $x=0$,
depends on the values for the local  curvature of the boundaries.  The
other two periodic orbits, indicated in the figure as dashed lines are
unstable for any set of the  billiard's parameters and thus, give rise
to a symmetric ternary horseshoe.  In Fig.~\ref{fig:knee-horseshoe} we
show this ternary horseshoe superimposed to the scattering layer given
in Birkhoff coordinates over the lower (parabolic) boundary.

\begin{figure}[!t]
\begin{center}
\includegraphics[scale=0.6]{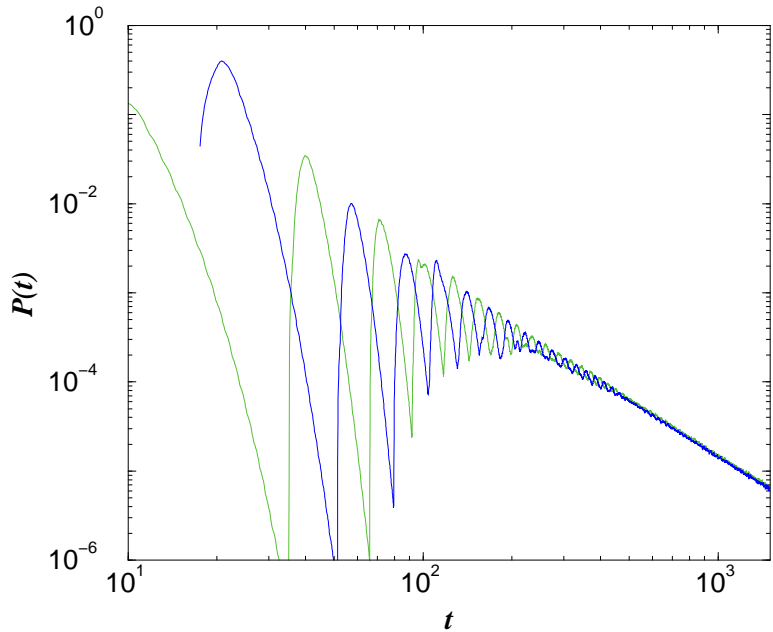}
\caption{\label{fig:knee-echoes}    Outgoing    flux   of    classical
trajectories as a function of  time for the knee billiard, measured at
$q=-20$ (blue) and $q=20$ (green).}
\end{center}
\end{figure}

The phase portrait of this billiard  shows a generic structure with an
underlying ternary horseshoe.  Therefore,  the appearance of echoes is
guaranteed. Again  we have  obtained the  outgoing flux  of  classical
trajectories  as a  function  of  time  $P(t)$, measured at  symmetric
positions   in   the    asymptotic  region.      This  is  shown    in
Fig.~\ref{fig:knee-echoes} in which the scattering echoes are evident.
As  before, the echoes  emitted to the left  are in counter phase with
those emitted to the right.

\section{Conclusions and outlook}
\label{sec:conclusions}

In a soft chaos  scattering situation,  where  a large  central island
exists, we find  a   self-pulsing  effect, which  we   call scattering
echoes.  They can be measured experimentally, and can  be used to make
significant   progress in  the inverse   scattering  problem for  such
situations.

We understand  the   inverse  scattering  problem   as the   task   to
reconstruct  information   about   the   chaotic invariant    set from
asymptotic data. The approach presented  is new in two aspects. First,
it   handles little developed  horseshoes  where  a homoclinic  tangle
encircles a large KAM island. Second, to our knowledge it is the first
method for  the  inverse chaotic  scattering  problem which  works for
classical and quantum ( wave ) dynamics along the same idea.

We concentrate  on systems described  by  binary and ternary symmetric
horseshoes in a two dimensional Poincar\'e  map. The generalization to
asymmetric  ternary  horseshoes is simple  and  we have  mentioned the
essential  equation   for this   case in eq.~\ref{eq:magic_3as}.  More
complicated generalizations would be necessary for billiards with more
than two openings.  Such systems can serve as models for rearrangement
scattering where each  opening represents one arrangement channel. The
corresponding Poincar\'e map has  a more complicated domain of several
connected components   which can lead   to horseshoes  of  essentially
complex structure.  The  treatment of  the   inverse problem  for such
systems   remains an  interesting  open  question. Another  nontrivial
generalization is    the  transition to systems  with   more  than two
important degrees of freedom. Yet the theory  of chaotic scattering in
higher dimensions is incipient \cite{wiggins}.

The test of our ideas in laboratory experiments is in progress for the
transmission  of    electromagnetic  waves  through   open    cavities
\cite{exp04}.  Such experiments   are typically done  in the frequency
domain, but if we use  Fourier transforms to  pass to the time domain,
the following fascinating  possibility exists: We  measure a very long
sequence of echoes.   The first ones  correspond to components of  the
wave,  which  only   visit  the  outer  regions   of   the  homoclinic
tangle.  After longer time we measure  components, which tunneled into
the  outer regions of the  stable island.   Signals leaving after ever
longer   times   correspond   to    tunneling deep     into  the   KAM
island.   Accordingly the change  of   the distances between  adjacent
echoes during  the measurement  reflects the differential  rotation of
the various layers inside of the  island.  In this type of measurement
the absorption which exists in any real experiment will set a limit to
the maximal time  (depth of penetration into  the island) which can be
reached. The problem of  absorption  might be circumvented by  placing
antennae into the interior of the  interaction region. Of course, then
the experiment is no longer a pure scattering experiment, and it would
no  longer be   a testing  ground for ideas   for  solving the inverse
problem. However, it might be a worthwhile experiment for the analysis
of tunneling of waves out  of KAM islands.   The only problem could be
the  distortion  of  the horseshoe   dynamics by the   presence of the
antenna.

While microwave billiards  present an obvious  testing ground  for our
results, we expect, that  scattering echoes of  the type we discussed,
will appear in other experiments in atomic  and molecular physics, and
that situations similar to the billiards can  be set up in mesoscopics
and for atoms in laser fields.

\ack
We  acknowledge  useful discussions    with F.  Borondo,  F.  Leyvraz,
P. Seba,  A. Heine, T. Friedrich,  A, Richter and  H.-J.  St\"ockmann.
Financial support  by  DGAPA-UNAM,  project IN101603 and   by CONACyT,
project 25-192-E is  acknowledged.  C.M.-M.  acknowledges a fellowship
by DGEP-UNAM and  the kind hospitality at  the Centro Internacional de
Ciencias (Cuernavaca) where part of this work was done.

\vspace{8mm}

\end{document}